\documentclass[twocolumn]{aastex631}

\usepackage{mathtools}
\usepackage{hyperref}

\DeclarePairedDelimiter\abs{\lvert}{\rvert}

\begin{document}

\title{JWST Reveals Powerful Feedback from Radio Jets in a Massive Galaxy at z = 4.1}

\correspondingauthor{Namrata Roy}
\email{nroy13@jhu.edu}

\author[0000-0002-4430-8846]{Namrata Roy}
\affiliation{Center for Astrophysical Sciences, Department of Physics and Astronomy, Johns Hopkins University, Baltimore, MD, 21218}

\author[0000-0001-6670-6370]{Timothy Heckman}
\affiliation{Center for Astrophysical Sciences, Department of Physics and Astronomy, Johns Hopkins University, Baltimore, MD, 21218}

\author[0000-0002-8214-7617]{Roderik Overzier}
\affiliation{Leiden Observatory, University of Leiden, Niels Bohrweg 2, 2333 CA Leiden, The Netherlands}
\affiliation{Observatório Nacional/MCTI, Rua General José Cristino 77, Rio de Janeiro, RJ 20921-400, Brazil}
\affiliation{TNO, Oude Waalsdorperweg 63, 2597 AK, Den Haag, The Netherlands}

\author[0000-0001-5333-9970]{Aayush Saxena}
\affiliation{Department of Physics, University of Oxford, Denys Wilkinson Building, Keble Road, Oxford OX1 3RH, UK}
\affiliation{Department of Physics and Astronomy, University College London, Gower Street, London WC1E 6BT, UK}

\author[0000-0001-6889-8388]{Kenneth Duncan}
\affiliation{Institute for Astronomy, University of Edinburgh Royal Observatory, Blackford Hill, Edinburgh, EH9 3HJ, UK}

\author[0000-0003-2884-7214]{George Miley}
\affiliation{Leiden Observatory, University of Leiden, Niels Bohrweg 2, 2333 CA Leiden, The Netherlands}

\author{Montserrat Villar Martín}
\affiliation{Centro de Astrobiología (CAB), CSIC-INTA, Ctra. de Ajalvir, km 4, 28850 Torrejón de Ardoz, Madrid, Spain}

\author[0000-0003-1020-1597]{Krisztina Éva Gabányi}
\affiliation{Department of Astronomy, Institute of Physics and Astronomy, ELTE E\"otv\"os Lor\'and University, P\'azm\'any P\'eter s\'et\'any 1/A, H-1117 Budapest, Hungary}
\affiliation{HUN-REN--ELTE Extragalactic Astrophysics Research Group, E\"otv\"os Lor\'and University, P\'azm\'any P\'eter s\'et\'any 1/A, H-1117 Budapest, Hungary}
\affiliation{Konkoly Observatory, HUN-REN Research Centre for Astronomy and Earth Sciences, Konkoly Thege Mikl\'os \'ut 15-17, H-1121 Budapest, Hungary}
\affiliation{CSFK, MTA Centre of Excellence, Konkoly Thege Mikl\'os \'ut 15-17, H-1121 Budapest, Hungary}

\author[0000-0001-5609-2774]{Catarina Aydar}
\affiliation{Max-Planck-Institut für Extraterrestrische Physik, Gießenbach-
straße, D-85748 Garching, Germany}

\author[0000-0001-8582-7012]{Sarah E. I. Bosman}
\affiliation{Institute for Theoretical Physics, Heidelberg University, Philosophenweg 12, D-69120, Heidelberg, Germany}
\affiliation{Max-Planck-Institut für Astronomie, Königstuhl 17, D-69117, Heidelberg, Germany}

\author[0000-0001-8887-2257]{Huub Rottgering}
\affiliation{Leiden Observatory, University of Leiden, Niels Bohrweg 2, 2333 CA Leiden, The Netherlands}

\author[0000-0001-8940-6768]{Laura Pentericci}
\affiliation{INAF – Osservatorio Astronomico di Roma, via Frascati 33, 00078, Monteporzio Catone, Italy}

\author[0000-0003-2984-6803]{Masafusa Onoue}
\affiliation{Kavli Institute for the Physics and Mathematics of the
Universe (Kavli IPMU, WPI), The University of Tokyo, 5-1-5 Kashiwanoha,
Kashiwa, Chiba 277-8583, Japan}
\affiliation{Center for Data-Driven Discovery, Kavli IPMU (WPI), UTIAS, The University of Tokyo, Kashiwa, Chiba 277-8583, Japan}
\affiliation{Kavli Institute for Astronomy and Astrophysics, Peking
University, Beijing 100871, P.R.China}

\author[0000-0002-6472-6711]{Victoria Reynaldi}
\affiliation{Instituto de Astrofísica de La Plata, CONICET–UNLP, CCT La Plata, Paseo del Bosque, B1900FWA La Plata, Argentina}
\affiliation{Facultad de Ciencias Astronómicas y Geofísicas, Universidad Nacional de La Plata, Paseo del Bosque, B1900FWA La Plata,
Argentina}



\begin{abstract}

We report observations of a powerful ionized gas outflow in a $z = 4.1$ luminous ($\rm L_{1.4GHz} \sim 10^{28.3} \ W \ Hz^{-1}$) radio galaxy TNJ1338-1942 hosting an obscured quasar using the Near Infrared Spectrograph (NIRSpec) on board JWST. We spatially resolve a large-scale ($\sim \rm $15 kpc) outflow and measure resolved outflow rates. The outflowing gas shows velocities exceeding 900 $\rm km \ s^{-1}$ and broad line profiles with line widths exceeding 1200 $\rm km \ s^{-1}$ located at $\sim$ 10 kpc projected distance from the central nucleus. The outflowing nebula spatially overlaps with the brightest radio lobe, indicating that the powerful radio jets are responsible for the extraordinary kinematics exhibited by the ionized gas. The ionized gas is possibly ionized by the central obscured quasar with a contribution from shocks. The spatially resolved mass outflow rate shows that the region with the broadest line profiles exhibits the strongest outflow rates, with an integrated mass outflow rate of $\sim$ 500 $\rm M_{\odot} \ yr^{-1}$. Our hypothesis is that an over-pressured shocked jet fluid expands laterally to create an expanding ellipsoidal “cocoon” that causes the surrounding gas to accelerate outwards. 
The total kinetic energy injected by the radio jet is about 3 orders of magnitude larger than the total kinetic energy measured in the outflowing ionized gas. This
implies that kinetic energy must be transferred inefficiently from the jets to the gas. The bulk of the deposited energy possibly lies in the form of hot ($\sim 10^7$ K) X-ray-emitting gas.

\end{abstract}



\section{Introduction} \label{sec:intro}

\subsection{Background}

Supermassive black holes (SMBHs), which are nearly ubiquitous in the nuclei of galaxies, play a crucial role in the evolution of galaxies. Large-scale cosmological simulations found that without incorporating energy released from active galactic nuclei (AGN),  massive galaxies overproduce young stars and fail to truncate star formation in time  \citep{dimatteo05, croton06}. Thus, they are unable to reproduce the galaxy population we observe today.  However, the detailed mechanism regarding this feedback process is still unknown. 

AGN can provide feedback to their surrounding environments through a variety of processes, including direct radiation \citep{ciotti10}, relativistic plasma jets \citep{fabian12}, and massive outflows of gas \citep{crenshaw03}. These energy outputs are sufficient to unbind or reheat most of the interstellar gas, if the energy is deposited efficiently in the ambient medium. 
In the context of galaxy evolution, there are two theoretically proposed feedback channels depending on the mode of energy transfer: winds driven as part of the bolometrically luminous ``quasar'' mode \citep{dimatteo05}, and the ``radio'' or the kinetic mode \citep{croton06, mcnamara07} where the primary means of energy transfer is via radio jets traced by emission from relativistically charged particles. Multiple observations have provided evidence of powerful high-velocity outflows in both these modes \citep{jarvis21, villar-martin21, cicone15, maiolino12, morganti21, speranza21}. Recently, \cite{heckman23} computed a global inventory of both feedback modes, and concluded that the radio jet mode is likely to be more energetically relevant.

Our detailed understanding of how radio jets impact the host galaxy interstellar medium (ISM) has made considerable progress in recent years using hydro-dynamical jet simulations \citep{wagner11, wagner12, mukherjee18, mukherjee20, dutta24}. These models show that collimated radio jets effectively deposit energy and momentum as they plough through the low-density channels of the ISM. Eventually, they inflate large cocoons/bubbles of shocked gas that can entrain the ambient gas leading to large-scale outflows and a kinematically disturbed ISM \citep{begelman89}. These predictions are in agreement with observations of nearby radio-AGN systems, which exhibit irregular ionized gas kinematics in the host galaxy ISM \citep{mukherjee16, wagner12, roy18, roy21, girdhar22, meenakshi22}. Similar observations of outflows driven by powerful radio galaxies at high-redshift (z$>$2) exist \citep{nesvadba17}, but are comparatively rare.

High-redshift radio galaxies (HzRGs) are excellent laboratories for studying AGN feedback during the late stages of massive galaxy evolution in the early Universe. They have high stellar masses, large star formation rates (exceeding a few hundred to often a thousand $\rm M_{\odot} \ yr^{-1}$), and host powerful radio jets. These jets are often spatially aligned with extended warm ionized gas nebulae spanning several tens of kiloparsec. They show evidence of increased velocity offsets, greater line widths, and disturbed kinematics in the host ISM of radio galaxies observed at $z>2$ \citep{villar-martin03, nesvadba06, nesvadba08, nesvadba17}.  Since these HzRGs host luminous obscured quasars (QSOs), the blinding glare of light from the central AGN is blocked, so it is straightforward to identify broad lines associated with outflowing gas in the galaxy ISM.

\begin{figure*}
    \centering
    \includegraphics[width=\textwidth]{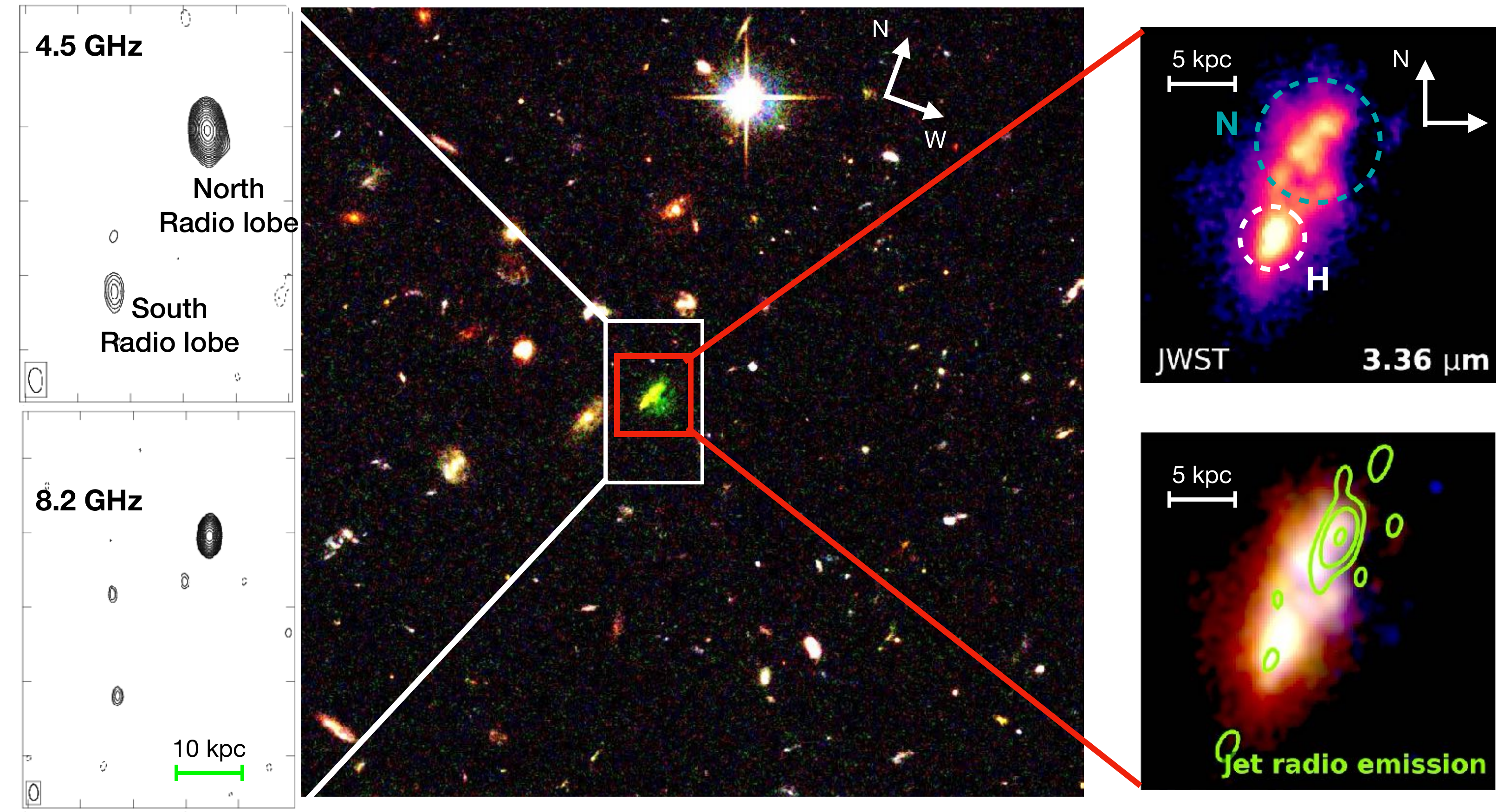}
    \caption{[Middle] Colour-composite image assembled from observations taken between July 8 and 12, 2002 by the HST ACS Wide Field Camera \citep{miley04}. The highlighted source in the center is our target radio galaxy TNJ1338, at z = 4.104, sitting in a protocluster environment. [Left] VLA radio continuum map at 4.5 GHz (top left) and at 8.2 GHz (bottom left), as reported in \cite{pentericci00}. TNJ1338 exhibits a clear double-lobed radio source with the northern lobe about $\sim$4 times brighter than the southern lobe. [Top right] JWST NIRCAM image taken with F335M filter around the northern radio lobe. The existing multi-band NIRCAM images \citep{duncan23} reveal two distinct components in TNJ1338: a compact host galaxy (H), with an extended clumpy emission-line dominated nebular (N) region spatially aligned with the bright northern radio lobe. [Bottom right] The RGB color image combines the ACS imaging for the blue channel, NIRCam/short wavelength imaging for the green channel, and NIRCam/long wavelength imaging for the red channel from \cite{duncan23}. Green contours show the VLA 8.46 GHz radio continuum emission associated with the galaxy, with contours starting at 4$\sigma$ (increasing by $\rm 4^n$, where n = 1, 2, 3, 4). The brightest northern radio lobe is spatially coincident with the nebular emission line-dominated region. } 
    \label{fig:tnj_RGB}
\end{figure*}

\subsection{Our target TNJ1338-1942}

The luminous high redshift radio galaxy TNJ1338-1942 \citep[hereafter TNJ1338;][]{debreuck99} is one of the most powerful radio sources known in the early Universe ($\rm L_{1.4\ GHz} \sim 10^{28.3} \ W \ Hz^{-1}$ at z = 4.104) and one of the first $z>4$ radio galaxy to be discovered in the southern sky. The radio morphology from VLA indicates a double-lobed structure, resembling a typical FRII radio galaxy, with a projected linear size of 36 kpc. The brightest radio lobe is located 8.8 kpc north of the host galaxy and the fainter one 27.7 kpc to the south (Fig.~\ref{fig:tnj_RGB}). The northern radio lobe is $\sim$ 4 times brighter than the southern component. The radio spectral index of this source is ultra-steep \citep[$\rm S_{\nu} \propto \nu^{-1.6}$;][]{pentericci00}, and the host galaxy resides at the core of a significant overdensity of galaxies, indicating the presence of a protocluster \citep{venemans02, miley04, zirm05, intema06, overzier08, overzier09, saito15}. TNJ1338 also has a 30 kpc-sized region of X-ray emission detected with Chandra X-ray observatory, coincident with the radio emission, which is suggested to be due to inverse Compton scattering of photons by relativistic electrons around the radio-AGN \citep{smail13}. Existing MUSE observation reveals an extremely luminous ($\rm L_{Ly\alpha} \sim 4 \times 10^{44} \ erg \ s^{-1}$) and spatially asymmetric giant Ly$\alpha$ halo \citep{debreuck99, venemans02, swinbank15}, which spans a massive scale of $\sim \rm 150 \ kpc$, and is spatially offset from the radio and X-ray emission. Also, this radio galaxy hosts a central obscured quasar detected in the infrared continuum \citep{falkendal19} and traced by very luminous optical emission lines. Although ground-based Near-IR IFU observations using VLT/SINFONI instrument do not spatially resolve this emission \citep{nesvadba17}, they show broad line profiles ($\sim 10^3$ $\rm km \ s^{-1}$), implying a powerful outflow \citep[see also ][]{debreuck99, swinbank15}.

This unique conglomeration of multi-wavelength properties in such a high redshift radio-loud source made TNJ1338 a prime target for JWST observations. TNJ1338 was observed with JWST NIRCAM as part of a Cycle 1 imaging program \citep[GTO 1176, PI: Windhorst;][]{windhorst23, duncan23} and also observed with JWST NIRSpec IFU to obtain spatially resolved rest optical spectroscopy \citep[GO 1964, PIs: Overzier and Saxena, ][]{saxena24}. Deep optical imaging from HST/ACS observations spatially resolved the optical emission tracing extended ionized gas in this galaxy. It revealed that the ionized gas structure is clumpy \citep{miley04, zirm05}. JWST NIRCAM images confirmed this finding, stating that there is indeed abundant extended ionized emission, and it consists of two distinct components -- one centered on the host galaxy center (marked as `H' in Fig.~\ref{fig:tnj_RGB} right), and the other nebular component roughly spatially coincident with the bright northern radio lobe \citep[marked as `N' in Fig.~\ref{fig:tnj_RGB} right;][]{duncan23}. The encounter of the northern jet with the denser gas in the north is a possible reason why the northern radio lobe is brighter and much closer to the host galaxy nucleus than the southern lobe.

\subsection{Questions to be investigated}

In this work, we utilize the NIRSpec IFU observations to study the spatially resolved ionized gas kinematics in the host galaxy (H) and nebular line-dominated region (N) to investigate how the radio jets interact with the ambient medium, and if the outflow energetics are sufficient to provide large-scale feedback in this high redshift radio source. A companion paper \citep{saxena24}  discusses the ionization structure of this galaxy in detail.


The impact of the outflow on the host galaxy ISM can be quantified by measuring the 
the outflowing mass (M), mass outflow rate ($\rm \dot{M}$), outflow velocity (v), momentum flow rate ($\rm \dot{p} = \dot{M} v$), and the kinetic power  ($\rm \dot{E} =  \frac{1}{2} \dot{M} v^2$) of the outflows. 
Simulations predict that efficient feedback with high coupling with the ISM needs a kinetic power of outflows to be at least 0.5-5\%  of the AGN bolometric luminosity, depending on the specific model \citep{dimatteo05, hopkins10}. Studies of NLR outflows in nearby Seyferts/ QSOs have measured a large range of kinetic efficiency $\rm \dot{E}/L_{bol}$ of $<0.1$\% to as high as 10\% for outflows in ionized phases \citep{baron19, fiore17, brusa16, oliveira21}, and an even larger range in mass outflow rates ($\rm 0.01 - 10^{3}  \ M_{\odot} \ yr^{-1}$). The majority of these studies have measured ``global'' outflow rates and power with single mass, velocity, and density estimates \citep{harrison14, mcelroy15, kakkad16, villar-martin16}. Many of these measurements are affected by seeing, and the outflow sizes and rates are massively overestimated.
Moreover, these integrated measurements result in an oversimplification of the kinematically complex systems and often result in up to 3 orders of magnitude uncertainties in the calculated AGN kinetic power \citep{villar-martin16, storchi-bergmann18, revalski18}. Spatially resolved observations are thus necessary to map out the spatial distribution of outflow velocities, densities, and ionized gas masses and to thus better constrain the outflow parameters as a function of spatial position \citep{revalski18, revalski21, crenshaw15, venturi18, kakkad22}.

We report here the first measurements of spatially resolved outflow rates and energetics from a $z>4$ radio galaxy using JWST, utilizing key rest frame optical features as ionized gas tracers.  We find that TNJ1338 shows spatially extended ionized gas with large-scale outflowing gas kinematics (outflow velocities $\rm  v_{outflow} \sim $ 800-1000 $\rm km \ s^{-1}$, line widths $\rm W_{80} $ exceeding 2000 $\rm km \ s^{-1}$) coincident with the bright radio lobe seen in Fig.~\ref{fig:tnj_RGB}. We will show that while the majority of the emission-line gas is probably photo-ionized by the central quasar ionizing radiation escaping along the radio jet axis, the extremely high-velocity gas is created due to the interaction of the radio jet with the environment.

The paper is organized as follows: Section 2 outlines the JWST NIRSpec data used in this study and the associated data reduction steps. Section 3 discusses the kinematic analyses performed on the reduced data cubes. The subsequent results are narrated in Section 4. In Section 5, we discuss the implications of the result and end with the conclusion in Section 6.

Throughout this paper, we assume a flat cosmological model with $H_{0} = 70$ km s$^{-1}$ Mpc$^{-1}$, $\Omega_{m} = 0.30$, and  $\Omega_{\Lambda} =0.70$, and all magnitudes are given in the AB magnitude system \citep{oke83}.

\section{Observations} \label{sec:obs}


\subsection{NIRSpec IFU observations} \label{subsec:data_nirspec}

The JWST/NIRSpec observations of TNJ1338 reported here were taken as part of the Cycle 1 General Observer (GO) program 1964 (PIs: Overzier and Saxena) on February 22nd, 2023 beginning at 22:43:58 UTC, using two grating-filter combinations: G235H/F170LP and G395H/F290LP. All the {\it JWST} data used in this paper can be found in MAST: \dataset[10.17909/v0kq-8381]{http://dx.doi.org/10.17909/v0kq-8381}. The NIRSpec IFU provides spatially resolved spectroscopy over a 3$'' \times$ 3$''$ field of view with 0.1$'' \times$ 0.1$''$ spatial elements.  
The observations were taken with an NRSIRS2 readout pattern and 16 groups per integration, with a total of 2 integrations in a 4-point dither pattern. An on-source integration time of $\sim$ 9.5 Ksec was used for each grating, with a total on-source exposure time of 18680 seconds. An equal integration time was spent to nod off-scene for good-quality background subtraction. The source was acquired using available deep HST and high-resolution VLA radio imaging. The resultant spectra cube has spectral resolution R $\sim$ 2700, over the wavelength range 1.66-5.27 $\mu$m (rest frame 3200 to 10300 \AA). 
For a full description of the data acquisition and reduction, we refer the reader to \cite{saxena24}.
Below we summarize briefly the key steps of the data reduction.

\subsection{NIRSpec data reduction} \label{subsec:data_reduce}

All the analyses and reduced data cubes presented in this paper are based on the Space Telescope JWST pipeline version 1.11.1 \citep{stsci}, and the context file "\texttt{jwst\_1118.pmap}" made available on August 2023. This latest calibration reference file incorporates proper on-orbit flat-fielding and improved flux calibrations. All the individual raw images are first processed for detector-level corrections using the \texttt{Detector1Pipeline} module of the pipeline (Stage 1). The input to this stage is raw non-destructively read ramps. A number of detector-level corrections are performed in this stage such as data quality and saturation check, dark current subtraction, super bias subtraction, fitting ramps of non-destructive group readouts, linearity, and persistence correction. The output is uncalibrated count-rate images per exposure per integration. 

The resultant images are then processed with \texttt{CALWEBB\_SPEC2}, Stage 2 of the pipeline.  
This step assigns a World Coordinate system (WCS) to each frame, flags pixels affected by MSA ``failed'' open shutters, and performs flat field correction. An image-from-image background subtraction is performed using the observations of the dedicated off-scene background. Finally, flux calibrations are performed with the most updated in-flight calibrations to convert the data from count rates to flux density in cgs units. We skip the imprint subtraction step since it increases the overall noise level in the final data cube. The individual Stage 2 images are resampled and co-added in the final Stage 3 of the pipeline (\texttt{CALWEBB\_SPEC3}). Before combining the individual data cubes, an outlier-detection step is required to identify and flag cosmic rays and other artifacts in the reduced data. However, the ``outlier detection'' step in the pipeline tends to identify an excess of false positives and occasionally marks strong emission lines as outliers. We therefore skip the default algorithm and use our custom routine that determines outliers across the different dither positions. We utilize a sigma-clipping technique to remove the outliers. We additionally used python package \texttt{astroscrappy} \citep{curtis18}, which is a python version of the widely used \texttt{LACosmic} routine \citep{vandokkum01}, to remove any other residual outliers missed by the sigma clipping technique. We mask the outliers and
exclude them from the data sets. The cleaned dithered observations are assembled to create the final 3D data cube using the \texttt{cube\_build} step. We select the Exponential Modified Shepard Method (``emsm'') of weighting when combining detector pixel fluxes to provide high signal-to-noise at the spaxel level. The datacube is produced with a spaxel size of 0.1”. 

The uncertainty in NIRSpec IFU astrometry can be anywhere between $\sim \rm 0.1-0.3''$. There is a spatial offset between the astrometry of NIRSpec IFU obtained from this program (GO 1964) and the NIRCAM images of TNJ1338 obtained from the PEARLS collaboration \citep[GTO 1176 and 2738,][]{duncan23} by $\sim$0.1$''$. This is likely dominated by the pointing uncertainty of the IFU mode. The NIRCAM mosaics are matched with Gaia-DR3 (after applying proper motion corrections) and are much more accurate, with an uncertainty level of $\rm \pm 0.02''$. Hence, we re-register our NIRSpec IFU maps by aligning the continuum map obtained from our NIRSpec observations (shown later in Fig.~\ref{fig:narrowband_images} top left panel) with the NIRCAM/F300M derived continuum image. Thus, the spatial alignment between our NIRSpec maps and the VLA radio continuum images (presented in Fig.~\ref{fig:tnj_RGB}, and shown later in Figs.~\ref{fig:narrowband_images}, \ref{fig:velchannel}, \ref{fig:o3moment}, \ref{fig:o3linewidth}) has an associated uncertainty $\rm \pm 0.02''$.


\section{Data Analyses} \label{sec:data_analyses}

\begin{figure*}
    \centering
    \includegraphics[width=\textwidth]{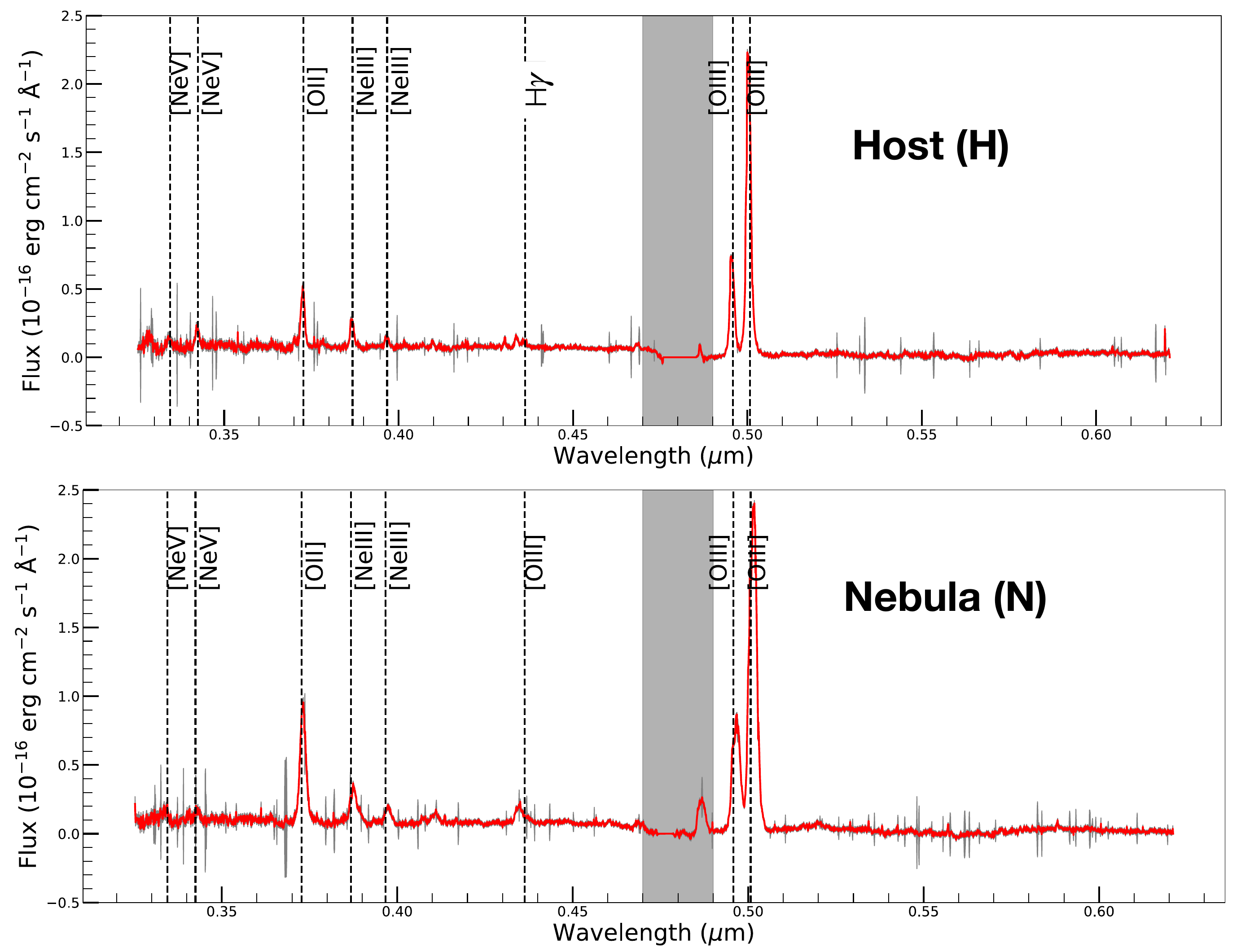}
    \caption{[Top panel] The JWST/NIRSpec IFU spectra extracted from the host galaxy region (H) in the southern part of TNJ1338. The spectra shown are within a central circular aperture of radius 2 kpc for the blue wavelength channel. The gray-shaded region shows 1-$\sigma$ errors in the measured fluxes. The rest-frame optical nebular emission lines visible in the spectral window are marked by black dashed lines. [Bottom panel] Spectra extracted from the nebular emission-line dominated region (N), within a circle of radius 4 kpc. The circular apertures encompassing the H and N regions are marked in Fig.~\ref{fig:tnj_RGB}.  } 
    \label{fig:host_nebular_spectra}
\end{figure*}

Spatially resolved rest-frame optical emission lines are obtained with the JWST NIRSpec IFU spectroscopy for our source TNJ1338. Data reduction pipeline steps, outlined in \S 2.2, produce wavelength and 
flux-calibrated, combined, and rectified spectra for each spaxel in a 3-dimensional data cube. Fig.~\ref{fig:host_nebular_spectra} shows the spectra taken with G235H grating, extracted within the host (H) and the nebular (N) apertures, as marked in Fig.~\ref{fig:tnj_RGB}. These spectra were obtained by summing the spectra from the spatial pixels lying within the respective apertures where the signal-to-noise of [OIII]$\lambda$5007 line exceeded 3$\sigma$. Multiple emission lines are detected, including  [OIII]4959\AA, [OIII]5007\AA, [OII]3727, 29\AA, H$\beta$, [NeIII]3868\AA, and [NeV]3435\AA \ -- and many others in G395H grating (H$\alpha$+[NII] 6584\AA, and [SII]6717, 31\AA). We derive the systemic redshift: z = 4.104$\pm$0.001, from the measured wavelength of the integrated [OIII]5007\AA \  emission extracted within the host galaxy (H marked in Fig.~\ref{fig:tnj_RGB}) as shown in Fig.\ref{fig:host_nebular_spectra}. This is roughly in agreement with the previously reported value of z = 4.1057 $\pm$ 0.0004, based on the HeII emission line from MUSE rest-ultraviolet  (UV) spectra \citep{swinbank15}.  The discrepancies of the order of $\sim$ 450 $\rm km \ s^{-1}$ possibly arise because the flux-weighted center of the HeII line used by \cite{swinbank15} could be significantly offset from that of the host position. Besides, the presence of chaotic kinematics and powerful outflows can affect all the prominent rest UV to near-infrared emission lines. Our JWST/NIRSpec measurements are able to spatially constrain the precise location of the host galaxy. Moreover, [OIII] is a more reliable estimator of the systemic redshift that rest UV lines.  Thus in our study, all the emission lines are constrained to have our measured z=4.104. 
 
Our analyses are primarily based on emission lines. So the first step is subtraction of the continuum level from the reduced data cubes per spaxel. To model the continuum emission around each emission line, we select spectral windows to that specific line's immediate red and blue side. For our most widely used [OIII]5007\AA \ emission line, we select continuum windows at rest frame 4900-4925\AA \  in the blue side and 5075-5100\AA \ in the red side.  For [OII]3727, 29\AA \ doublet, we select 3660-3680\AA \ and 3778-3880\AA, and for H$\beta$, we choose 4820-4840\AA \ and 4880-4900\AA. These wavelength ranges are selected to be closest to the said line while ensuring that there is no contamination from any emission line. Next, we fit a second-order polynomial curve within the wavelength range bounded by the above wavelength window using \texttt{SCIPY}'s least square optimization module. The fitted continuum is subtracted from the emission line spectra, and the process is repeated for every spaxel to create a continuum-subtracted emission line cube. This method provides a robust subtraction of the continuum level for each line.

\begin{figure*}
    \centering
    \includegraphics[width=\textwidth]{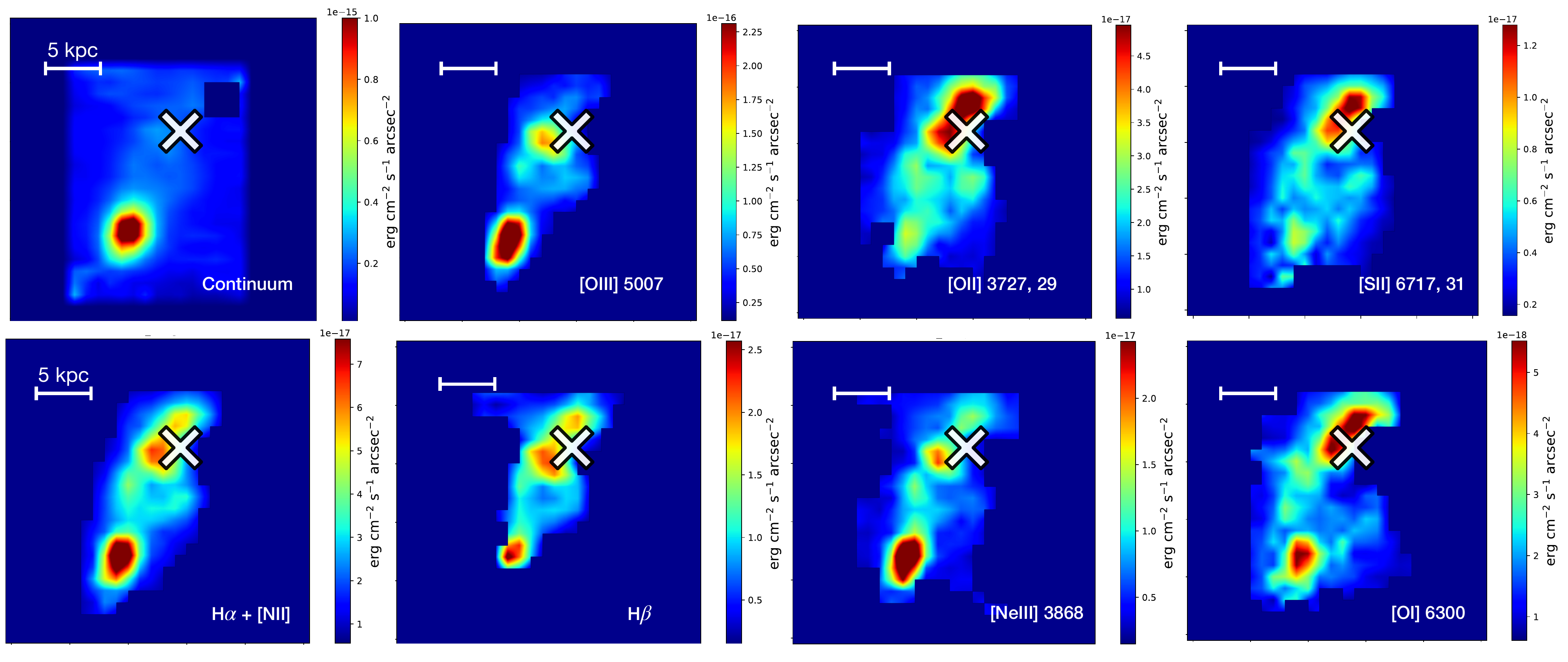}
    \caption{We present line-integrated continuum subtracted flux maps of [OIII] 5007 \AA, [OII] 3727, 29 \AA, [SII] 6717, 6731 \AA, H$\alpha$ + [NII] 6584\AA, H$\beta$, [NeIII] 3868 \AA, and [OI] 6300 \AA \ in the different panels, in units of $\rm erg \ s^{-1} \ cm^{-2} \   arcsec^{-1}$.   The maps show only the spaxels with line flux detected at S/N$>$3.  
 The first panel [top left] shows the continuum map, where all emission lines are masked. North is vertical. The host galaxy (H) in the southern end of our source is represented by the strong continuum presence (marked in Fig.~\ref{fig:tnj_RGB}). The northern region is dominated by nebular (N) emission-line gas and exhibits diverse morphology in different line tracers. This extended emission region is spatially aligned (and roughly coincident) with the brightest radio lobe of the radio galaxy (marked by a white cross). The extended nebula is very prominent in the low ionization lines ([OI], [OII], [SII]) compared to the host gas, and thus indicates an ionization level that is quite low compared with the host gas. More details are in \cite{saxena24}. }
    \label{fig:narrowband_images}
\end{figure*}

We constructed pseudo-narrow-band (NB) images for the detected emission lines using the continuum-subtracted NIRSpec datacube. The maps are centered on the position and wavelength of the corresponding line. We used a wide spatial aperture to include all the detectable emissions associated with TNJ1338, above the noise level. The spectral bandwidths for constructing the NB image were chosen to include more than 95\% of the total integrated line flux and to maximize the signal-to-noise within the circular aperture of radius 2 kpc (0.3$''$) centered on the host galaxy. 
The resulting continuum-subtracted NB images for seven different emission lines are shown in Fig.\ref{fig:narrowband_images}.

We create moment 0 (intensity), moment 1 (velocity), and moment 2 (dispersion) maps from the [OIII]$\lambda$ 5007 emission line data cubes after the continuum level has been robustly subtracted out. Any spatial pixel with emission line signal-to-noise $<$ 2 is masked out. This threshold is chosen to be able to capture the faint emission structures while simultaneously discarding the noisy regions beyond our object of interest.  
We also non-parametrically calculate the different tracers of gas velocity dispersions -- e.g. full-width half maxima (FWHM), W50, and W80 for those two emission lines, without performing any model fitting. This enables us to directly extract the kinematics from the data and not make any assumption about the shape of the emission line (i.e. whether it is Gaussian, Lorentzian, Voigt profile, etc).
   
$\rm W_{50}$ is defined as the line width containing 50\% of the emission line flux, and $\rm W_{80}$ as the line width containing 80\% of the emission line flux. The W50 map of an emission line is calculated by computing the cumulative distribution function (CDF (x) = P(X $\leq$ x), which is the probability that the random variable X is less than or equal to x) of the normalized line flux values for each spaxel. 
The velocity values corresponding to 25\% ($\rm v_{25}$) and 75\% ($\rm v_{75}$) of the integrated flux values are calculated from the CDF evaluated at x = 0.25 and 0.75. Then, $\rm W_{50}$ = $\rm v_{75} - v_{25}$. Similarly, $\rm W_{80}$ is calculated from the difference of $\rm v_{90}$ and $\rm v_{10}$. We calculate the FWHM by finding half of the peak flux on either side of the line profile, and  
calculating the distance between those points.


\section{Results} \label{sec:results}
\subsection{Emission-Line Morphology} 

The morphology of the warm ionized gas in our sources traced by the different emission lines are
shown in Fig.~\ref{fig:narrowband_images}. The emission maps of all the lines are very irregularly shaped and show extended nebulosities of ionized gas. In an upcoming paper, \cite{saxena24} discuss the ionization structure and the detailed morphology and emission line ratios of the different ionized gas tracers. The size of the extended ionized region is roughly $\sim$ 15 kpc along the major axis and $\sim$ 7 kpc in the perpendicular direction.  The rest-frame optical continuum, shown on the first panel, is not contaminated by line emission. The continuum map is obtained by collapsing over the line-free wavelength region over the full available spectral bandwidth. Our source has a single, unresolved continuum source indicating the location of the host galaxy with an obscured quasar. 

[OIII] 5007\AA \  is the strongest emission line in our NIRSpec data cube, and thus the best tracer of ionized gas and outflows in our target TNJ1338. We will primarily focus on the [OIII] emission in the rest of this section. Fig.~\ref{fig:narrowband_images} (second panel) shows the continuum subtracted narrowband [OIII] image, produced by the method described in \S\ref{sec:data_analyses}. Two distinct [OIII] bright regions are clearly visible. The peak in the [OIII] surface brightness map coincides with the optical-continuum-bright region and hence indicates the emission from the host galaxy (marked as `H' in Fig.~\ref{fig:tnj_RGB}) with an obscured quasar.  However,  nearly all the rest of the [OIII] line-emitting gas lying in the upper part of the map indicates a predominantly ``emission-line-dominated'' region (region `N' in Fig.~\ref{fig:tnj_RGB}) with almost no detectable continuum with SN $>$5.  
This extended plume of [OIII]  is roughly aligned with the radio jet axis and spatially coincides with the bright northern radio lobe (shown in Fig.~\ref{fig:narrowband_images} using a white cross). The line emission extends to at least 12 kpc from the center of the host galaxy.  The excellent angular resolution ($\sim \rm 0.1''$) and sensitivity of JWST NIRSpec reveal these intricate sub-structures within the ionized gas distribution.

As we will discuss in \S\ref{subsec:o3_kin}, the measured large velocities, increased line widths, and asymmetric line profiles measured in the extended ionized gas are clear outflow signatures. Their spatial coincidence with the radio lobe possibly indicates a scenario where the jet encounters dense gas and inflates a cocoon/ bubble, which entrains gas clouds 
creating these observed outflows.

\begin{figure*} 
    \centering
    \includegraphics[width=\textwidth]{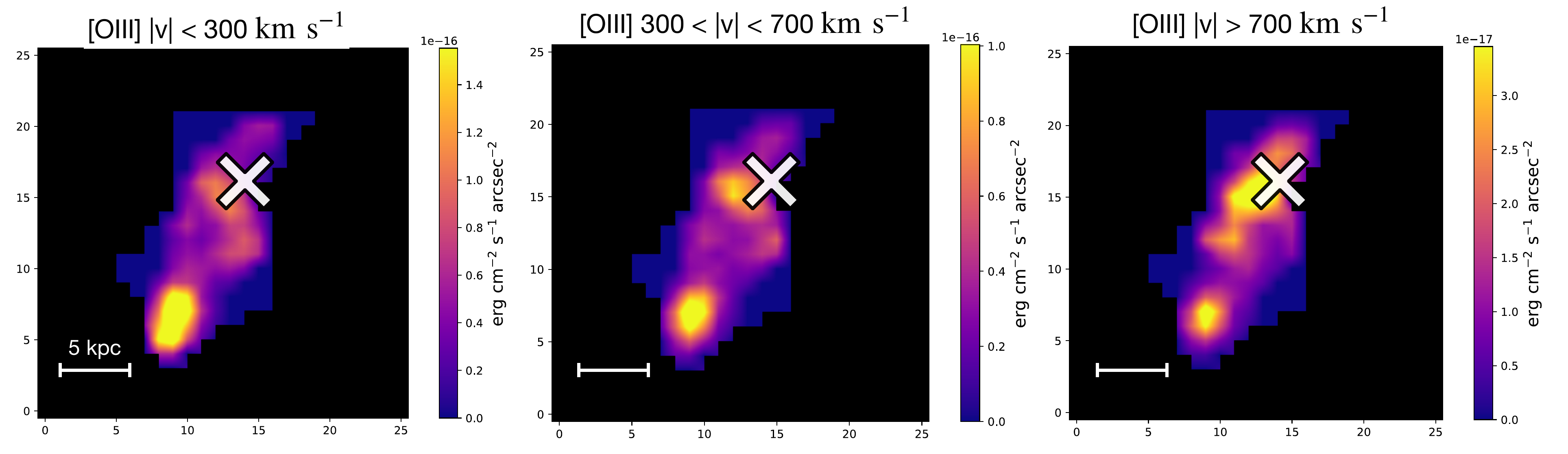}
    \caption{Velocity channel maps for the [OIII] 5007 \AA\ line emission. The left panel shows the [OIII] line flux with velocities within $\pm$ 300 $\rm km \ s^{-1}$ around the systemic velocity ($\rm v_{sys}$), thus highlighting the slowest moving ionized gas. The majority of the [OIII] emitting gas with a velocity close to $\rm v_{sys}$ lie near the host galaxy region in the southern part of TNJ1338. The middle panel shows the intermediate velocity components and the right panel highlights the fastest-moving part of the ionized wind with v $>$ + 700 $\rm km \ s^{-1}$ and v $< -$ 700 $\rm km \ s^{-1}$. The white cross shows the spatial position of the bright radio lobe from the VLA image.  It is clear that the fastest-moving gas coincides with the nebular region where the bright radio lobe lies. Our hypothesis is that an expanding ``cocoon'' driven by the shocked jet fluid accelerates the ambient gas outwards to large velocities, which may be perceived as blueshifted and redshifted gas on either side of the galaxy (see \S\ref{subsec:jet-feedback}). All the maps show only the spaxels with S/N$>$3 on the total [OIII] flux.  }
    \label{fig:velchannel}
\end{figure*}

\subsection{[OIII] kinematics} \label{subsec:o3_kin}

Fig.~\ref{fig:velchannel} shows the [OIII] 5007 emission maps in three different velocity channels: [-300, 300] $\rm km \ s^{-1}$,  [$\abs{300}$, \  $\abs{700}$] $\rm km \ s^{-1}$ (which includes [-300, -700] and [300, 700] $\rm km s^{-1}$), and [$\abs{700}$, \ $\abs{1500}$] $\rm km~s^{-1}$.  

We see similar patterns in H$\alpha$+[NII], and [OII] lines, but as mentioned before, we restrict our kinematic discussion primarily to [OIII] 5007 emission since this is a bright forbidden line and is an ideal tracer of outflows. A collimated emission with a linear morphology is evident most prominently in the highest velocity bin ( $\abs{v} >$ 700 $\rm km \ s^{-1}$). Indeed, the majority of the rapidly moving gas emission is located in the nebular line-dominated region and is roughly coincident with the spatial location of the radio lobe detected in early VLA data. The [OIII] emission in this velocity bin is peaking at a projected distance of $\sim$ 9 kpc from the peak of the continuum emission from the host galaxy. The velocities we measure in this region exceed 1000 $\rm km \ s^{-1}$ and are well in excess of gas undergoing rotational motion under the influence of gravity. This morphology clearly confirms a fast-moving [OIII] outflow, as previously hinted in \cite{debreuck99, swinbank15}. We also detect some high-velocity gas within the inner few kpc of the central region of the host galaxy. This can be driven by the central quasar. However, in this work, we mostly focus on the large-scale outflow in the nebular [OIII] dominated region.

As we move to the lower velocity bins ($\abs{v} =$ 300 to 700 $\rm km \ s^{-1}$), a remarkable change in the [OIII] morphology is apparent. The brightest emission in this velocity range now peaks at the host galaxy centered on the continuum, but a filament of line emission becomes evident at the [OIII]-emitting region -- extending perpendicular to the fastest moving [OIII] outflow probed in the highest velocity bin. This intermediate velocity emission seems to connect the lowest velocity [OIII] ($\abs{v} <$ 300 $\rm km \ s^{-1}$) centered on the host galaxy to the high-velocity outflowing gas lying coincident with the radio lobe.

\begin{figure*}
    \centering
    \includegraphics[width=\textwidth]{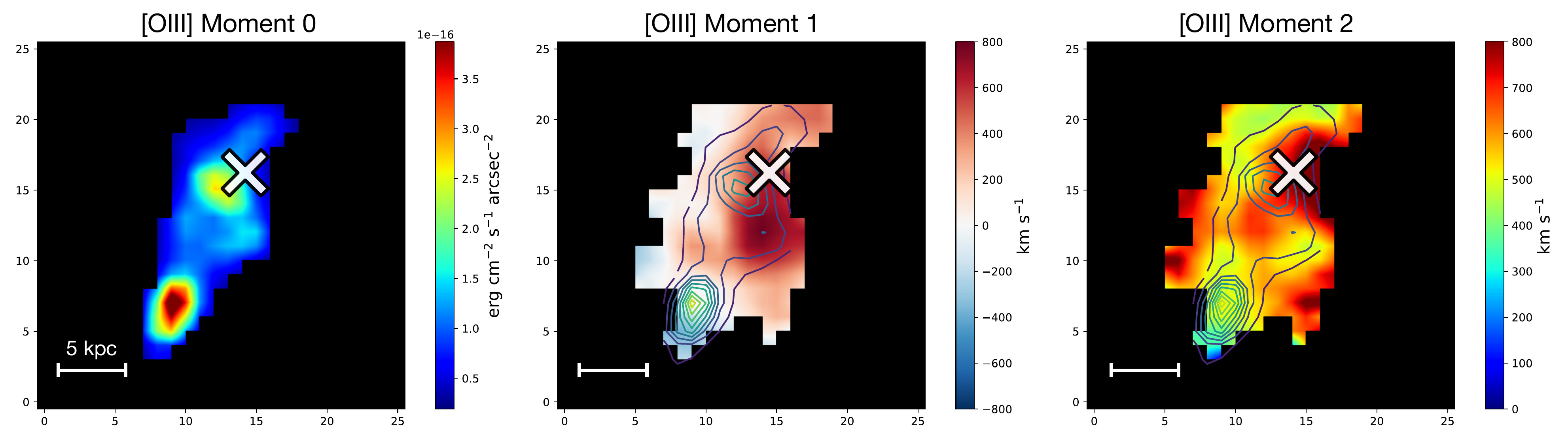}
    \caption{Moment 0 [left panel], Moment 1 [middle panel], and Moment 2 [right panel] maps of the [OIII] 5007 \AA \ line emission of TNJ1338. The black contours in the middle and right panel indicate the [OIII] moment 0 contours. The white cross marks the position of the bright radio lobe. The host galaxy exhibits an intermediate velocity ($\rm |v| \sim 400 \ km \ s^{-1}$), while the nebular region exhibits velocity exceeding 1000 $\rm km \ s^{-1}$. The velocity shifts (moment 1) are measured with respect to the systemic velocity. The presence of blue-shifted and red-shifted velocity components indicates an expanding bubble/cocoon, possibly driven by the shocked radio jet, which in turn entrain and accelerate gas outwards in radial directions. The moment 2 values are very high in the spatial locations coincident with the radio lobe, indicating broad line widths ($
    \sigma >$ 1000 $\rm km \ s^{-1}$) and turbulent motions in the ambient medium where the bow shock is generated at the terminus of the jet. See \S\ref{subsec:jet-feedback} for a more detailed discussion.    }
    \label{fig:o3moment}
\end{figure*}

The kinematics maps in Fig.~\ref{fig:o3moment} also strongly highlight the features we discussed above. This figure shows the Moment 0, Moment 1, and Moment 2 maps of the total [OIII]$\lambda$5007 emission with Moment 0 contours overplotted on top. The Moment 1 map shows that the highest velocity offset from the systemic, i.e. v $>$ 1000 $\rm km \ s^{-1}$, is indeed exhibited by the emission-line-dominated region associated with the northern radio lobe (shown by white cross). The host galaxy region shows an absolute velocity offset lower than 400 $\rm km \ s^{-1}$. The fainter, filamentary structure with intermediate velocity ( 300  $< \abs{v} <$ 700 $\rm km \ s^{-1}$ ) connects the host galaxy AGN with the outflow region. 
The absolute Moment 1 values depends on the exact determination of the zero-point, or redshift of the system. Owing to the chaotic gas kinematics of this system, we derive the redshift from the measured [OIII] velocity at the central stellar continuum peak. The presence of blue-shifted and red-shifted velocity components indicates a laterally expanding bubble/cocoon of emission line gas, possibly driven by the
shocked radio jet, which may accelerate gas outwards in radial directions. See later in \S\ref{subsec:jet-feedback} for a more detailed discussion of the jet-driven feedback scenario. 

These large velocity values and the distinct gas morphology at different velocity channels are incompatible with the rotational motion associated with the host galaxy. We also see an extremely high gas dispersion exceeding 1200 $\rm km \ s^{-1}$ coincident with the [OIII] emitting gas associated with the largest velocity. Fig.~\ref{fig:o3linewidth} shows the non-parametric characterization of the width of the emission lines, measured using the process discussed in  \S\ref{sec:data_analyses}, and show very similar results. We show $\rm W_{50}$, $\rm W_{80}$, and FWHM for the [OIII] 5007 line. The maximal line width is not associated with the central regions of the host, as probed by
the position of the continuum peak. Using all three tracers of emission line widths, we find extremely broad line widths ($\rm W_{50} > $ 1200 $\rm km \ s^{-1}$, $\rm W_{80} > $ 2000 $\rm km \ s^{-1}$) spatially associated with the northern radio lobe, consistent with the velocity dispersion map. The increased line widths provide clear evidence that the extreme gas kinematics must be an outflow. The spatial coincidence with the radio lobe implies that this is likely a signature of a jet-driven outflow. The host galaxy shows a moderate line width with a mean $\rm FWHM_{avg} \sim $ 800 $\rm km \ s^{-1}$ within the central 0.5$''$ around the peak of the stellar continuum.  
This is consistent with our interpretation from the highest velocity channel maps in Fig.~\ref{fig:o3moment} that the upper [OIII] enhanced region must be associated with high-velocity outflowing gas clouds. 
 
In Fig.~\ref{fig:o3profile}, we show the [OIII]5007 spectral profiles at six spatial locations (aperture size = 0.1$''$) along the straight line connecting the host galaxy and the high-velocity outflow region.  Although the host galaxy also shows a clear signature of a broad outflow component in the line profile, the spectral components become broader as we move toward the emission-line-dominated region. There are two distinct components clearly visible with each component having FWHM $>$ 700 $\rm km \ s^{-1}$. We avoid spectral fitting throughout our analyses, because of these extremely broad spectral profiles where deblending into different components becomes very uncertain due to parameter degeneracy.  However, in this step, we want to extract the peak-to-peak separation between the two distinct components. Hence, we fit the spectral profiles using two Gaussian components to extract the location of the two peaks, without worrying about the exact number of kinematic components needed to optimally fit the spectra.  We plot the peak separation ($\rm v_{sep}$) and find that $\rm v_{sep}$ also exceeds 1200 $\rm km \ s^{-1}$ in agreement with $\rm W_{50}$ and FWHM estimates. Thus, to summarize, we identify:

\begin{figure*}
    \centering
    \includegraphics[width=\textwidth]{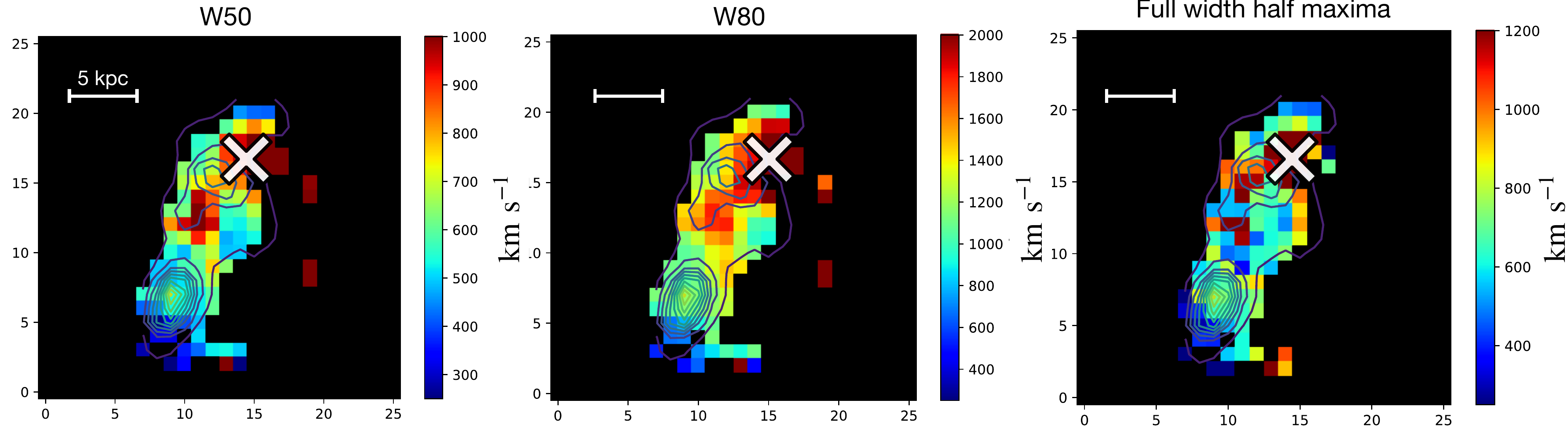}
    \caption{Different kinematic tracers (line widths) of the [OIII] line emission. [Left panel] Map of $\rm W_{50}$ = $\rm v_{75} - v_{25}$, which is the line width containing 50\% of the emission line flux. Here $\rm v_{75}$ and $\rm v_{25}$ are velocities at the 75th and 25th percentile of the overall emission line profile in each spaxel. [Middle panel] Map of $\rm W_{80}$, where $\rm W_{80} = v_{90} - v_{10}$. [Right panel] Full-width half maxima of [OIII] line profiles. See \S\ref{sec:data_analyses} for a detailed description of computing these non-parameteric line widths. Solid contours in all three panels show [OIII] flux, the white cross indicates the brightest radio lobe. Broad line widths are prevalent in the emission line-dominated nebular region (marked as `N' in Fig.~\ref{fig:tnj_RGB}), coincident with the radio lobe. $\rm W_{50}$ and FWHM exceeds 1200 $\rm km \ s^{-1}$, where $\rm W_{80}$ exceeds 2000 $\rm km \ s^{-1}$. This is consistent with Fig.~\ref{fig:o3moment}.  }
    \label{fig:o3linewidth}
\end{figure*}

\begin{table*}[ht]
    \centering
    \begin{tabular}{||l| c | c | c | c | c | c | c | c | c | c ||}
    \hline \hline
      z  & ${\rm M_{\star}}$ &  ${\rm L_{[OIII]}}$  &  ${\rm L_{jet}}$ & ${\rm L_{AGN}}$ & ${\rm M_{gas}}$   & ${\rm \dot{M}_{out}}$ & ${\rm KE_{gas}}$  &  ${\rm \dot{KE}_{outflow}}$ &  ${\rm \dot{P}_{outflow}}$  &  LAS  \\
       & ${\rm M_{\odot}}$ & ${\rm erg \ s^{-1}}$ & ${\rm erg \ s^{-1}}$ & ${\rm erg \ s^{-1}}$ & ${\rm M_{\odot}}$ & ${\rm M_{\odot} \ yr^{-1}}$ & erg & ${\rm erg \ s^{-1}}$ & dyne & kpc \\ \hline 
       
4.104 & $\rm 10^{10.9}$ & 5.6$\times \rm 10^{45} $ & 1.4$\times \rm 10^{47} $ & 2.15$\times \rm 10^{48} $ & 4.8 $\times 10^{9}$  & 497 & 2.7$\times \rm 10^{58}$ & 1.01$\times \rm 10^{44} $ & 2.06$\times \rm 10^{36} $ & 36 \\

\hline \hline
    \end{tabular}
    \caption{\textit{Measured global properties of TNJ1338. All quantities reported here are integrated over the whole galaxy. The columns are redshift, stellar mass, extinction corrected [OIII] 5007 \AA \ luminosity, radio jet mechanical energy derived from 1.4 GHz radio luminosity, AGN bolometric luminosity derived from dust corrected [OIII], ionized gas mass, mass outflow rate, kinetic energy of the outflow, kinetic power of the outflow, momentum rate, and the radio linear size respectively. The ionized gas mass, outflow rates, kinetic energy, and momentum rate are all derived per spaxel primarily using [OIII] 5007 \AA \ emission in a spatially resolved manner, as shown in Fig.~\ref{fig:energetics}. The values reported in this table are obtained by spatially integrating these resolved estimates over the spaxels where the [OIII] S/N $>$ 3 (per spaxel). Note, the $\rm \dot{M}_{out} $, $\rm KE_{gas}$, and $\rm \dot{KE}_{outflow} $ values reported here do not include corrections for projection effects (See \S \ref{subsec:o3_energy}) and hence are lower limits. } }
    \label{tab:source}
\end{table*}

\begin{table*}[ht]
    \centering
    \begin{tabular}{||l| c | c | c | c | c | c | c | c | c | c ||}
    \hline \hline
       Location  &  $\rm L_{[OIII]}$ & $\rm L_{H\alpha}$ & $\rm n_e$ &  $\rm \bar{W}_{50}$ &  $\rm \bar{W}_{80}$ & $\rm \Delta_{peak}$ & $\rm M_{gas}$ & $\rm \dot{M}_{out}$ & $\rm KE_{gas}$ & $\rm \dot{KE}_{outflow} $ \\
         &  $\rm erg \ s^{-1}$ & $\rm erg \ s^{-1}$ & $\rm cm^{-3}$ &  $\rm km \ s^{-1}$ &  $\rm km \ s^{-1}$ & $\rm km \ s^{-1}$ & $\rm M_{\odot}$ & $\rm M_{\odot} \ yr^{-1} $ & erg & $\rm erg \ s^{-1} $ \\ \hline 

        Host & $\rm 2.39 \times 10^{45}$  & $\rm 7.37 \times 10^{44}$  & 570 & 489  &  1064 & 125 & 2.8 $\times 10^{9}$ & 234 & 6$\times \rm 10^{57}$ & 2.05$\times \rm 10^{43}$ \\
        Nebula & $\rm 3.18 \times 10^{45}$ & $\rm 7.73 \times 10^{44}$ & 610 & 778  & 1668  &  748 & 1.9 $\times 10^{9}$& 264 & 2$\times \rm 10^{58}$ & 8$\times \rm 10^{43}$ \\
     \hline \hline
    \end{tabular}
    \caption{\textit{Measured properties at the host (H) and nebular (N) region of the TNJ1338, averaged/ integrated over the spatial apertures marked as `H' and `N' in Fig.~\ref{fig:tnj_RGB}. The columns are -- location in the galaxy (whether host or nebular), extinction corrected [OIII] 5007 \AA \ luminosity integrated over the host/nebular region,  extinction corrected H$\alpha$ luminosity, mean electron density, mean [OIII] $\rm W_{50}$ (line width containing 50\% of the emission line flux), mean $\rm W_{80}$ (line width containing 80\% of the emission line flux), the peak-to-peak separation between the two fitted kinematic components of the [OIII] line profile calculated per spaxel (Fig.~\ref{fig:o3profile}) and averaged over the host/nebular apertures, ionized gas mass, mass outflow rate, kinetic energy and kinetic power of outflows. The gas mass, outflow rates, and outflow energy are all computed per spaxel and integrated over spatial apertures bounding the host and nebular region. Note, the $\rm \dot{M}_{out} $, $\rm KE_{gas}$, and $\rm \dot{KE}_{outflow} $ values reported here do not include corrections for projection effects (See \S \ref{subsec:o3_energy}) and hence are lower limits. } }
    \label{tab:nebular}
\end{table*}

\begin{enumerate}
    \item A bright [OIII]-emitting region with a high velocity offset ($\rm \abs{v} > $ 700 $\rm km \ s^{-1}$) and broad line width (FWHM $>$ 1200 $\rm km \ s^{-1}$) with a bubble-like morphology. This kinematic component is centered at a projected separation of $\rm \sim 0.9''$ from the center of the host galaxy's continuum emission. This fast-moving component spatially coincides with the brighter radio lobe of the central radio galaxy and indicates a jet-driven outflow scenario. 

    \item A region with comparatively narrow (FWHM $\sim$ 700 $\rm km \ s^{-1}$) emission-lines centered close to systemic velocity ($\rm \abs{V} < $ 300 $\rm km \ s^{-1}$). This is associated with the host galaxy which harbors a central obscured quasar. 

    \item A region with intermediate velocity offsets (300 $ < \rm \abs{V} < $ 700 $\rm km \ s^{-1}$) and line widths (600 $<$FWHM$<$ 1000 $\rm km \ s^{-1}$) connecting the host galaxy center with the outflow. 
    
\end{enumerate}

\begin{figure*}
    \centering
    \includegraphics[width=\textwidth]{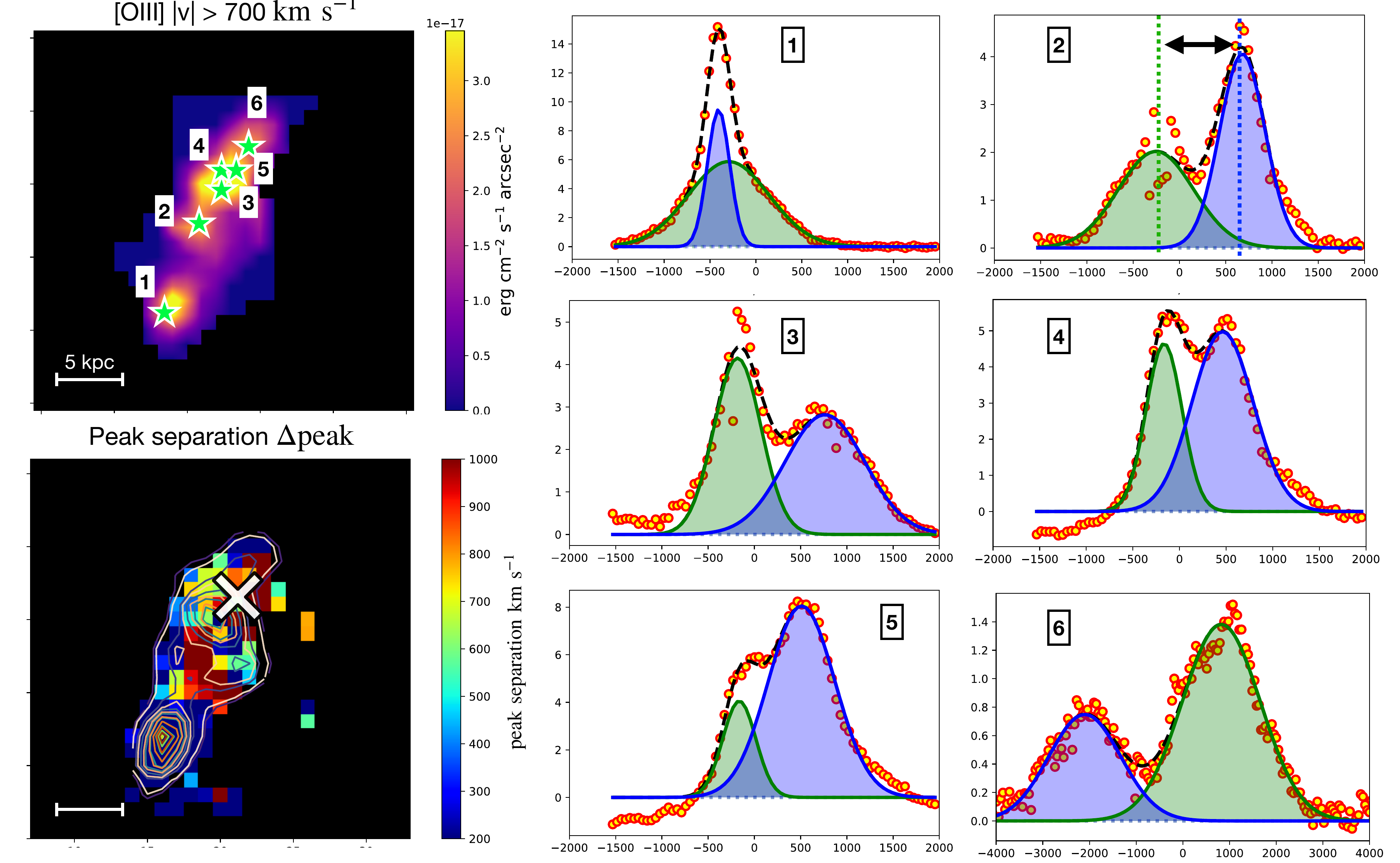}
    \caption{[Top left] The channel map for [OIII] 5007 \AA \ line, with $\rm |v| > 700 \ km \ s^{-1}$. [Right panel] Emission line profile around [OIII] 5007 emission line shown for six different spatial positions, marked by green stars in the top left panel. The line profiles indicate multi-component velocity profiles, with broad lines and extended `winged' line shapes. Clearly, the emission line profiles become more kinematically complex as we approach the nebular region, close to the radio lobe. We fit the line profiles with two components and determine the `peak-to-peak' separation between the two components. This provides a conservative estimate of the velocity offset of the [OIII] gas from the large-scale `bulk' motion of the gas. The line profiles may, in reality, require more kinematic components for an optimum fit. [Bottom left] The resolved map of peak-to-peak separation ($\rm \Delta peak$) of [OIII] line profiles, extracted from each spaxel, using the line profile fitting technique mentioned above. The white cross indicates the spatial location of the bright radio lobe, as before. $\rm \Delta peak$ exceeds $\rm 1000 \ km \ s^{-1}$ in the northern nebular region, consistent with Fig.~\ref{fig:o3moment} and ~\ref{fig:o3linewidth}.
    }
    \label{fig:o3profile}
\end{figure*}

Similar broad and fast-moving gas is detected in other strong emission lines as well -- particularly H$\alpha$, [OII], [NII] and [SII]. As shown in the narrowband images in Fig.~\ref{fig:narrowband_images}, bright, linear, extended emission is indeed coincident with the radio lobe (white cross) observed at the outflowing region in all the emission maps, very similar to the [OIII] emission, albeit with some small-scale structural differences. We leave the detailed analyses of the kinematics of the other strong emission lines and the similarities between each other for future work.

\subsection{Outflow rates \& energetics} \label{subsec:o3_energy}

In the previous section, we presented clear detection of fast-outflowing gas, traced by  the 
broad [OIII]$\lambda$5007 emission lines  (FWHM $>$ 700 $\rm km \ s^{-1}$ in Fig.~\ref{fig:o3linewidth}) and high velocity offset ($\rm \abs{v} > $ 1000 $\rm km \ s^{-1}$) across a large part of our target galaxy TNJ1338.  In this Section, we will study the spatial distribution of the mass of the outflowing gas, outflow rates, and the associated energetics of the warm (T $\sim \rm \ 10^{4}$ K) ionized gas. The extremely high spatial resolution ($\sim 0.1''$) and sensitivity of the JWST/NIRSpec IFU data presented here enables a detailed investigation of the ionized outflow rates and energetics across the entire galaxy. We note that this is the first time such a high-resolution characterization of a jet-driven outflow has been done in a $ z > 4 $  radio galaxy hosting an obscured quasar. Previous IFU studies at similar redshifts \citep{nesvadba17} were ground-based data with an angular resolution of $\sim$ 0.5 to 1 arcsec. There have been several other ground-based resolved studies of extended outflow in high redshift radio galaxies \citep{van97, villar-martin03, humphrey07}, but the lack of sufficient angular resolution has affected the energetics measurements. The majority of the studies 
have relied on spatially unresolved information and the detection of broad absorption features of UV resonant lines \citep{wang18, onoue19, bischetti22, bischetti23}. These features primarily trace nuclear outflows which encompass a small fraction of the galactic scale outflow. Moreover, the absence of spatial information prevents a reliable estimate of the outflow parameters. In the section below,
we outline how we estimate the mass, outflow rate, and energetics of the ionized outflow, and discuss the potential mechanisms responsible for driving the galaxy scale outflow.

\begin{figure}
    \centering
    \includegraphics[width=0.47\textwidth]{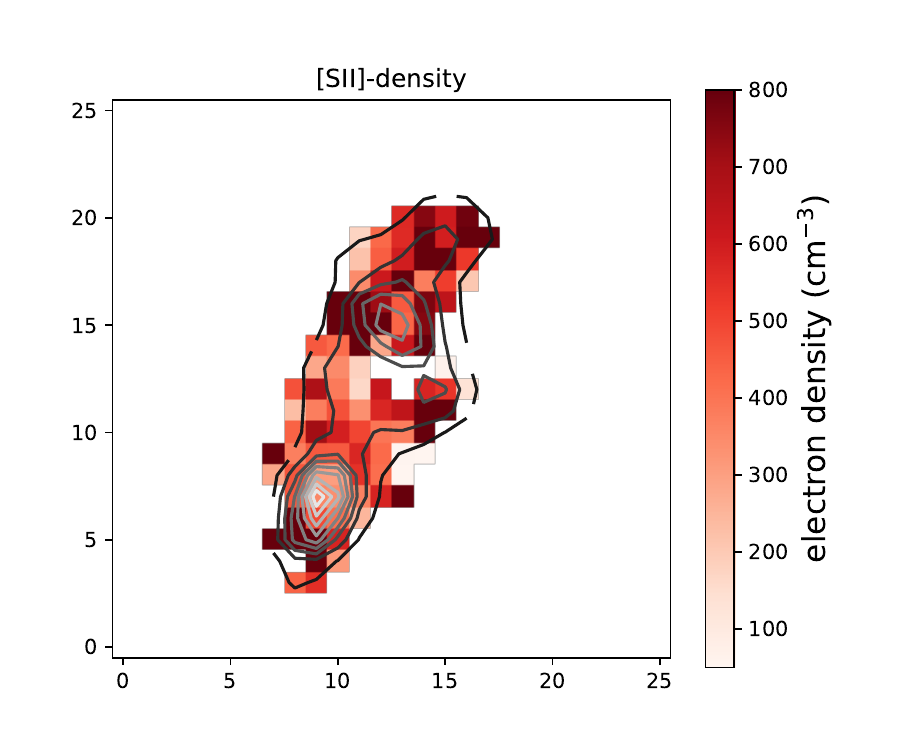}
    \caption{The electron density map of TNJ1338, derived using the line ratios of the [SII] 6717, 6731 \AA \ doublet ratio (see \S\ref{subsec:o3_energy}). The solid contours represent the [OIII] emission flux. The density ranges from 200-1000 $\rm cm^{-3}$, with mean values of 570 $\rm cm^{-3}$ and 610 $\rm cm^{-3}$ in the host and nebular region respectively (see Table.~\ref{tab:nebular}). }
    \label{fig:density}
\end{figure}

The first step is an accurate measurement of electron density $\rm n_e$ and its spatial variation at various locations within the galaxy. We derive $\rm n_e$ using the [SII]$\lambda \lambda$6717,30 line ratio. Each of the two [SII] lines was fitted using a two-component Gaussian for every spaxel, and the two components are tied together by the same velocity and dispersion value obtained from the brighter [OIII] 5007 moment 1 and 2 maps. This step is necessary to ensure accurate density measurements by extracting the line ratios properly. We wrap the fitting module into the python package \texttt{dynesty} \citep{speagle20, koposov22}, which uses a dynamic nested sampling algorithm \citep{skilling04, skilling06} to execute a multi-parameter optimization and extract Bayesian posteriors and evidence. Although there might still be unaccounted uncertainty owing to the simple double-gaussian model we fit, we strictly avoid overfitting the complex line profiles. We used the total [SII] $\lambda\lambda$ 6716,30 lines flux, the parametrization of \cite{sanders16}, and the assumption of $\rm T_e$=$\rm 10^4$ K to convert to electron density. In Fig.~\ref{fig:density}, we show the resulting electron density map. The density varies between 200 - 1100 $\rm cm^{-3}$, and the nebular emission-dominated region with the fast outflow signatures shows the highest values of the density. This is consistent with other local and high-$z$ AGN studies, which have found higher density estimates in the outflow component compared to the rotating disks in the host galaxy \citep{villar-martin14, mingozzi19, cresci23}.

\begin{figure}
    \centering
    \includegraphics[width=0.47\textwidth]{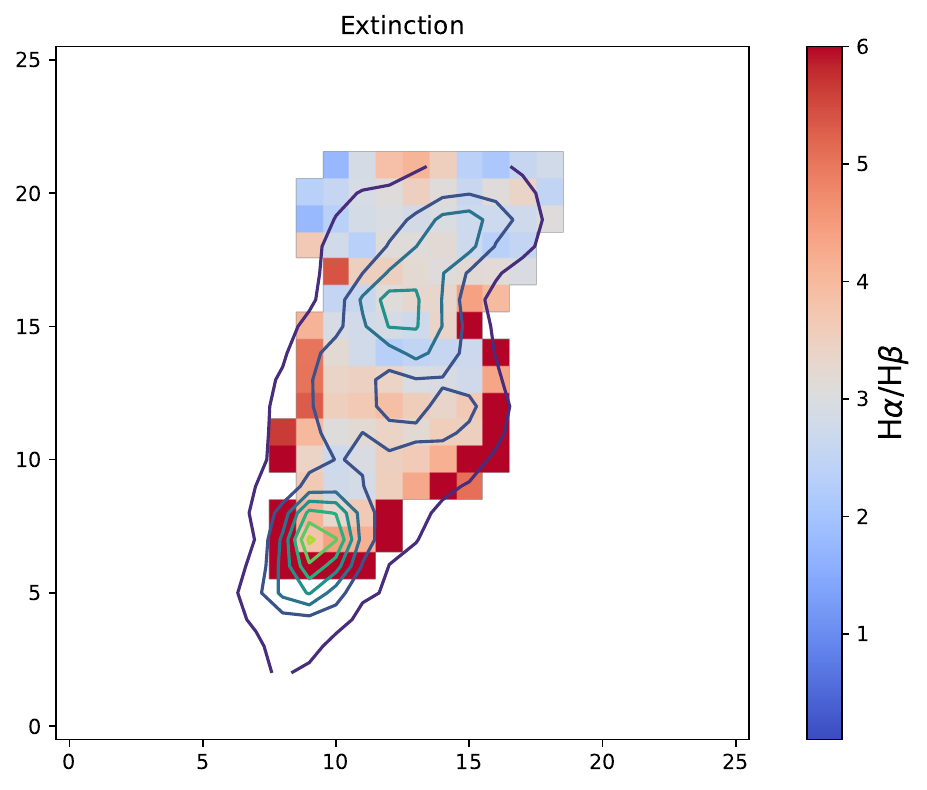}
    \caption{Balmer decrement measurement, i.e. H$\alpha$ / H$\beta$ emission line ratios, with [OIII] 5007 flux overplotted as contours. These measurements are used for dust correcting the [OIII] luminosities to compute ionized gas mass and outflow rates. H$\alpha$/H$\beta$ ratio spans from 7.11 in the host galaxy to 3.53 in the nebular region. }
    \label{fig:extinction}
\end{figure}

Next, we calculate the Balmer decrement i.e. H$\alpha$/H$\beta$ as shown in Fig.~\ref{fig:extinction}, which is a direct probe of dust extinction. Indeed, the host galaxy of TNJ1338 harbors an obscured quasar in the center and is thus expected to be dust-enshrouded with a higher H$\alpha$/H$\beta$ ratio. We indeed find that the average H$\alpha$/H$\beta$ is the highest in the central region of the host galaxy, with a median value = 4.29$\pm$1.1. For the extended gas, the ratio goes down to = 3.1 $\pm$0.8. 
The extinction estimates are derived from the measured Balmer decrement, assuming a standard R = 3.1 extinction curve \citep{osterbrock06}, and the temperature of the ionized gas to be $\sim \rm 10^4 \ K$. The mean extinction measured $\rm A_{v}$ = 0.8. We perform dust correction for the emission line fluxes before computing the outflow rates and energetics.

\begin{figure*}
    \centering
    \includegraphics[width=\textwidth]{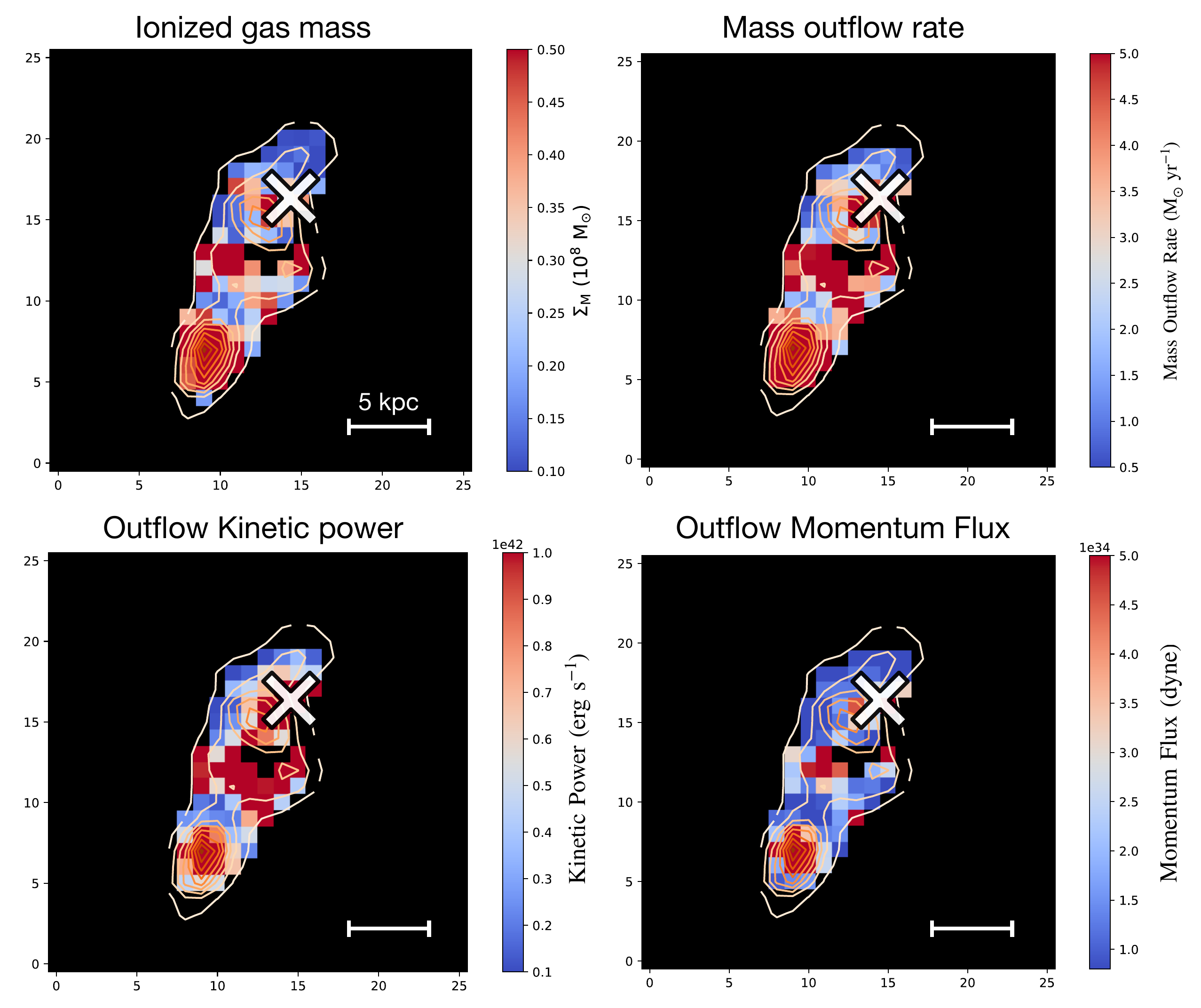}
    \caption{ [Top left] Spatially resolved map of ionized gas mass surface density, in units of $\rm 10^{8} \ M_{\odot}$ per spaxel. The ionized gas mass is derived from extinction-corrected [OIII] luminosity and [SII]-derived electron density (Eqn.~\ref{eqn:mion}). The total ionized gas mass integrated over the whole galaxy is 4.8 $\times \rm 10^{9} \ M_{\odot}$ (Table ~\ref{tab:source}). The solid contours represent [OIII] 5007 \AA \ emission flux and the white cross shows the location of the bright radio lobe. [Top right] Resolved map of ionized gas mass outflow rate, in units of $\rm M_{\odot} \ yr^{-1}$. Mass outflow rates are enhanced in discrete locations in both the host and the nebular regions, but with comparative values when integrated over spatial apertures marked as H and N (234 $\rm M_{\odot} \ yr^{-1}$ and 264 $\rm M_{\odot} \ yr^{-1}$; Table ~\ref{tab:nebular}). [Bottom left and Bottom right]  Outflow kinetic power and momentum rate in units of $\rm erg \ s^{-1}$ and dyne respectively. The total kinetic power of the outflow $\rm \dot{KE}_{outflow} \sim \ 10^{44} \ erg \ s^{-1}$, which is a small fraction of the kinetic power of the radio jet (Table \ref{tab:source}), implying that the transfer of the kinetic power from the jet to the gas is not very efficient. See detailed discussion in \S\ref{subsec:jet-feedback}.     }
    \label{fig:energetics}
\end{figure*}

In this work, we have discussed large-scale outflow signatures using the [OIII]$\lambda$5007  flux distribution, velocity, and line widths. The [O III]$\lambda$5007 emission line is the brightest
line emitted by the outflow within the rest-frame visible range covered by the NIRSpec data cube. It is also
well separated in wavelength from neighboring emission lines. 
Hence, we use dust-corrected [OIII]$\lambda$5007 to calculate the ionized gas mass and energetics using the formula below \citep{cano-diaz12, veilleux20, veilleux23}:

\begin{equation} \label{eqn:mion}
   \rm  M_{ionized} = 5.3\times 10^8 \frac{C_e L_{44}([OIII]\lambda 5007)}{n_{e,2}10^{[O/H]}} M_{\odot}
\end{equation}

where $\rm L_{44}([OIII]\lambda 5007)$ is the luminosity of [OIII] $\lambda$5007, normalized to $\rm 10^{44} \ erg\ s^{-1}$, $\rm n_{e,2}$ is the electron density normalized to $\rm 10^2 cm^{-3}$, $C_e$ is the electron density clumping factor which can be assumed to be of order unity, and $\rm 10^{[O/H]}$ is the oxygen-to-hydrogen abundance ratio relative to solar abundance. We assume solar abundance, so [O/H] = 0, and thus $\rm 10^{[O/H]}$ is 1. We use the spatially resolved density values derived from the [SII] 6716, 6731 flux ratio as shown in Fig.~\ref{fig:density}. The median electron density $\rm n_e = 570 \ cm^{-3}$, but it goes up to $\sim \rm 1150 \ cm^{-3}$  in the outflowing regions. We use [OIII] 5007 line fluxes corresponding to the velocity channel $\rm \abs{v} > 500 \ km \ s^{-1}$  to include only the outflowing gas component that is well in excess of observed velocities due to orbital motions (Maximum velocity consistent with orbital motion due to gravitationally bound orbits for TNJ1338 $\sim$ 250 $\rm \ km \ s^{-1}$). We corrected this flux for extinction using the H$\alpha$/H$\beta$ line ratio (see Fig.~\ref{fig:extinction}) and converted it to luminosity, which yielded a total luminosity $\rm L_{[OIII],corr} = 5.6 \times 10^{45} \ erg \ s^{-1}$. Note, for some of the spaxels, the H$\beta$ emission line falls in the detector chip gap. The dust correction for those spaxels are made by taking the median H$\alpha$/H$\beta$ line ratio, centered on the host galaxy.
In Fig.~\ref{fig:energetics} (upper left panel), we show the distribution of ionized gas obtained associated with the outflowing component. The extinction-corrected ionized mass surface density ranges from $\rm 2 \times 10^7  - 5 \times 10^7 \ M_{\odot}$ per spatial pixel, with a total outflowing mass $\rm M_{ionized} = 4.8 \times 10^{9} \ M_{\odot}$ (Table 1). The amount of ionized gas mass in the host vs the nebular region is tabulated in Table 2.

An estimate of ionized gas mass can also be derived using the outflow component of H$\beta$ or H$\alpha$ recombination line \citep{perna23, marshall23, cresci23}. 
To demonstrate this, we also measured dust-corrected H$\beta$ emission from the outflow, which yielded a total outflowing mass $\rm M_{ionized\_H_{\beta}} = 3.8 \times 10^{9} \ M_{\odot}$. This is consistent with what we obtain from [OIII] 5007. However, we note here that for a handful of spatial pixels, the H$\beta$ emission line fell onto the NIRSpec detector ``chip-gap'' and thus returned no value. Hence the calculated $\rm M_{ionized\_H_{\beta}}$ is a lower limit. We return to [OIII] 5007 derived mass estimate for the outflow mass rate and energetics calculation described below.

Once the outflow mass is inferred, the spatially resolved ionized outflow rate for each parcel of gas within the outflowing regions of TNJ1338 can be calculated using the following equation :

\begin{equation}
    \rm \dot{M_{out}} = \frac{M_{out} v_{out}}{R_{out}}
\end{equation}

Here we assume that $\rm M_{out}$ is the ionized gas mass surface density, i.e. mass in each spatial pixel, as derived from equation \ref{eqn:mion}. We presume that each of these gas parcels with mass $\rm M_{out}$ is at a projected radial distance $\rm R_{out}$ from the host galaxy center, defined by the peak of the stellar continuum.   We take the velocity in each pixel $\rm v_{out} = \sqrt{W_{50}^2 + \Delta v^2}$, where 
 $\rm W_{50}$ is the linewidth corresponding to 50\% of the emission line flux (see \S\ref{subsec:o3_kin} for details), and $\Delta v$ correspond to the velocity offset from the systemic velocity. 
 We obtain a total mass outflow rate $\rm \dot{M_{out}} = 497 \pm 47 \ M_{\odot}/yr $, and the spatially resolved mass outflow rate map is shown in Fig.~\ref{fig:energetics} (upper right panel). This large amount of outflowing gas makes this source stand out as one of the earliest known radio galaxies with the strongest outflow signature. The corresponding kinetic power of the outflow is given by 

\begin{equation}
    \rm \dot{KE}_{outflow}  = \frac{1}{2} \times \dot{M_{out}} v_{out}^2
\end{equation}

We obtained a total kinetic power of 1.01$\times \rm 10^{44} \ erg \ s^{-1}$, and the total momentum rate $\rm \dot{p_{kin}} $ = $\rm \dot{M_{out}}$ $\rm v_{out}$ = $\rm 2.06 \times 10^{36} $ dyne. The resolved maps of these quantities are shown in Fig.~\ref{fig:energetics} (bottom row), and their values in the host galaxy H and the nebular region N are mentioned in Table 1. The detailed implications of these findings are discussed in \S\ref{sec:discussion}.

It is important here to discuss the impact of projection effects on these estimates. These will affect both $v_{out}$ and $R_{out}$. The morphology and kinematics of the gas in the northern nebular region imply that both the structure of the gas and its outflow direction are aligned with the axis of the radio jet. Let us then designate the angle between the radio jet axis and our line-of-sight as $\phi$. In this case the true value ($v_{out,t}$) of the outflow velocity {\it vs.} the projected value ($v_{out,p}$) along our line-of-sight is $v_{out,t} = v_{out,p}/cos \phi$, and the actual distance of the material from the nucleus {\it vs.} the projected distance is given by $R_t = R_p/sin \phi$. From this is clear that the above calculations of the total KE and momentum in the gas will be strict lower limits. Since we do not know $\phi$ we will just assume its median value of 60$^o$ (that is, half of 4 $\pi$ lies in the region with $\phi < 60^o$). In this case, the median value for $v_{out,t}$ would be 2 $v_{out,p}$, the median value for $R_{out,t}$ would be 1.15 $R_p$, and the median value for the true crossing time would be 0.58 $R_p/v_{out,p}$, The net effects of this are that the true $\rm KE_t$ is 4 times $\rm KE_p$, $\rm p_t = 2 \rm p_p$, and the true outflow rates for $\rm \dot{M}$, $\rm \dot{p}$, and $\rm \dot{KE}$ will be larger than the projected values by factors of 1.7, 3.5, and 6.9 respectively.

\section{Discussion} \label{sec:discussion}

Our analyses of the NIRSpec data cube confirm the presence of spatially extended fast outflow in TNJ1338. The new data reveal prominent, highly complex, filamentary structures that extend from the host galaxy northward for a projected angular distance of $\sim \rm  1.66''$ or $\sim \rm 12 \ kpc$. We detect high-velocity gas clouds ($\rm \abs{v} > 800 \ km \ s^{-1}$) marked by broad lines ($\rm W_{50} \sim 1000 \ km \ s^{-1}$) and extended emission tracing a bubble-like morphology. This is aligned with the radio lobe and indicates jet-driven feedback. Interestingly, we detect some high-velocity gas centered in the host galaxy as well (Fig.~\ref{fig:velchannel} right panel) which are spatially compact and tend to be quite concentrated within the inner few kpc. This is possibly driven by radiative feedback from the central obscured quasar in TNJ1338. The mechanical feedback driven by the radio jet, however, affects the gas kinematics to larger distances and over a much larger volume. 
In this section, we revisit the emission line luminosity, outflow kinematics, and energetics to discuss the source of the ionized gas and the driving mechanisms of the jet-driven outflow.

\subsection{Photoionization from the central quasar?}

\begin{figure}
    \centering
    \includegraphics[width=0.48\textwidth]{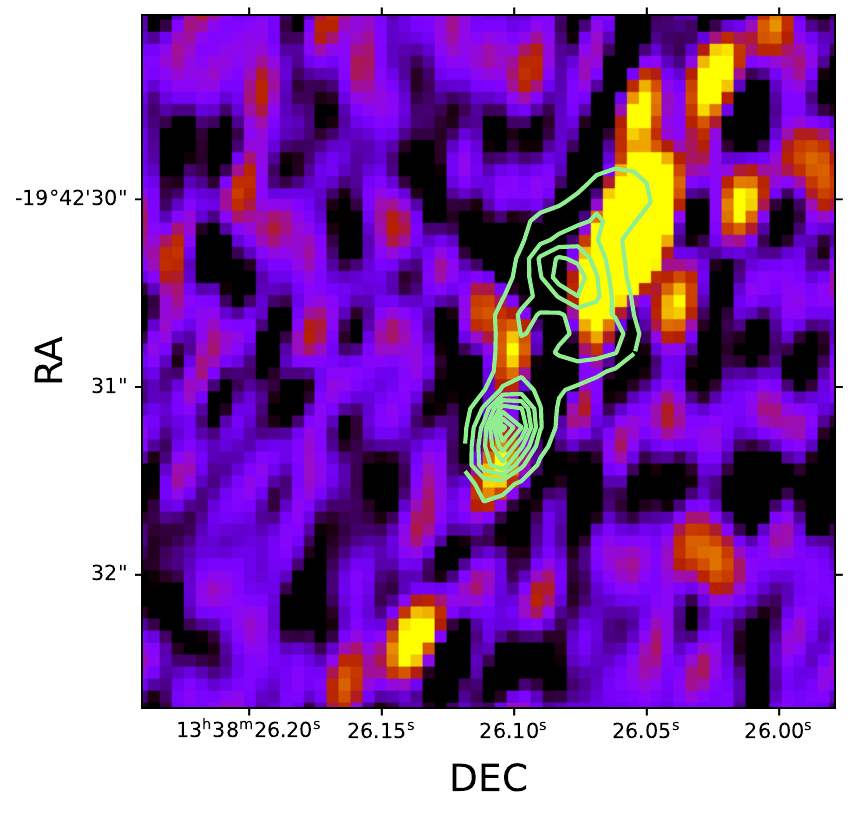}
    \caption{VLA 8.2 GHz radio continuum emission map with the [OIII] 5007 contours from JWST/NIRSpec overplotted on top. The radio galaxy shows a bi-polar jet with the brightest radio lobe aligned with the [OIII] enhanced region N. There can be an astrometric inaccuracy of the order of $\sim$ 0.02$''$ between the VLA and JWST/NIRSpec maps.  }
    \label{fig:radio_op_overlay}
\end{figure}

The first natural question is what ionizes the massive amount of $\sim \rm 5 \times 10^{9} \ M_{\odot}$ gas extending for tens of kpc from the nuclear region of the host galaxy. The most plausible explanation is that the [OIII]-emitting nebula is part of the interstellar or circumgalactic medium of the host galaxy TNJ1338, which is photoionized by the obscured quasar residing in the host galaxy and get ejected outwards by the large-scale outflow. We will return to the discussion about the nature and the source of the outflow in \S \ref{subsec:jet-feedback}; here we explore whether the central quasar is the primary source of ionization in this galaxy. 

The rest-frame ultraviolet continuum from a QSO is typically more than two orders of magnitude more luminous than what would be required to explain the typical observed [OIII] luminosities of the nebulae \citep{elvis94}.
Hence, photoionization by the quasar is unavoidable. To check if similar photoionization is occurring in  TNJ1338, we assume a model in which the nebulae are in photoionization
equilibrium with the diffuse QSO radiation field \citep[similar to ][]{crawford88, baum89, heckman91}. The electron number density of a photoionized cloud
($\rm n_e$) located a distance r from an ionizing source with a photon luminosity (Q) and characterized by an ionization parameter U (defined as the ratio of ionizing photons to electrons
in the cloud) is given by:

\begin{equation}
    \rm n_e = \frac{Q}{4\pi r^2Uc}
\end{equation}

 Our measured values of $\rm n_e$ range between 400 - 1200 $\rm cm^{-3}$, with a median value of 
570 $\rm cm^{-3}$. Value for U is inferred from the comparison of the relative
strengths of high-ionization and low-ionization lines to photoionization
models. We use dust corrected [OIII] $\lambda$5007/[OII]$\lambda\lambda$3726,3729 line ratio with assumed solar metallicity to determine a median ionization parameter U $\sim \rm 10^{-2} $.  Assuming r $\approx$ 10 kpc, yields Q = 1.9$\times\rm 10^{57} \ s^{-1}$. Assuming a mean spectral energy distribution of a quasar \citep{elvis94}, the relation between AGN bolometric luminosity ($\rm L_{AGN}$) and the ionizing luminosity (Q): $\rm L_{AGN}/Q = 9.5 \times 10^{-11} \ erg$ translates to a $\rm L_{AGN} = 1.8\times 10^{47} \ erg \ s^{-1}$. This is somewhat smaller than the $\rm L_{AGN} =2.15 \times  10^{48}$ $\rm erg \ s^{-1}$ derived from the far-IR and [OIII] 5007 luminosities (see Table 1). Thus, the central quasar seems more than capable of being the primary ionizing source, provided that a sufficient fraction of the ionizing radiation can escape the nucleus. The strong alignment of the extended emission-line gas and the radio source (
 Fig.~\ref{fig:radio_op_overlay}) implies that this gas is photoionized by the radiation that escapes the central quasar in a radiation cone that is aligned along the ``jet-axis''.  
We should note however that we can not entirely exclude ionization by shocks driven by the radio jet or its cocoon.

\begin{figure}
    \centering
    \includegraphics[width=0.45\textwidth]{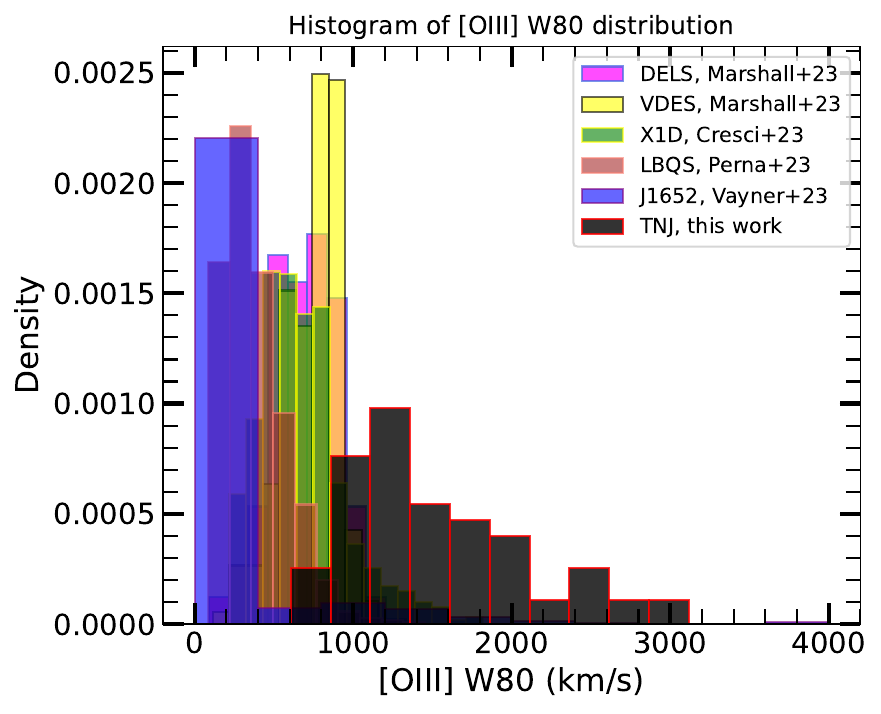}
    \caption{Histogram of the distribution of $\rm W_{80}$ values in radio-quiet quasars observed with JWST/NIRSpec, compared to our radio-loud galaxy TNJ1338 (black). Our source shows the highest line widths and the most extreme kinematics in the emission line gas. }
    \label{fig:kin_hist}
\end{figure}

\subsection{Comparison of TNJ1338 with radio-quiet Type 1 QSOs studied with JWST/NIRSpec IFU}

\begin{figure*}
    \centering
    \includegraphics[width=\textwidth]{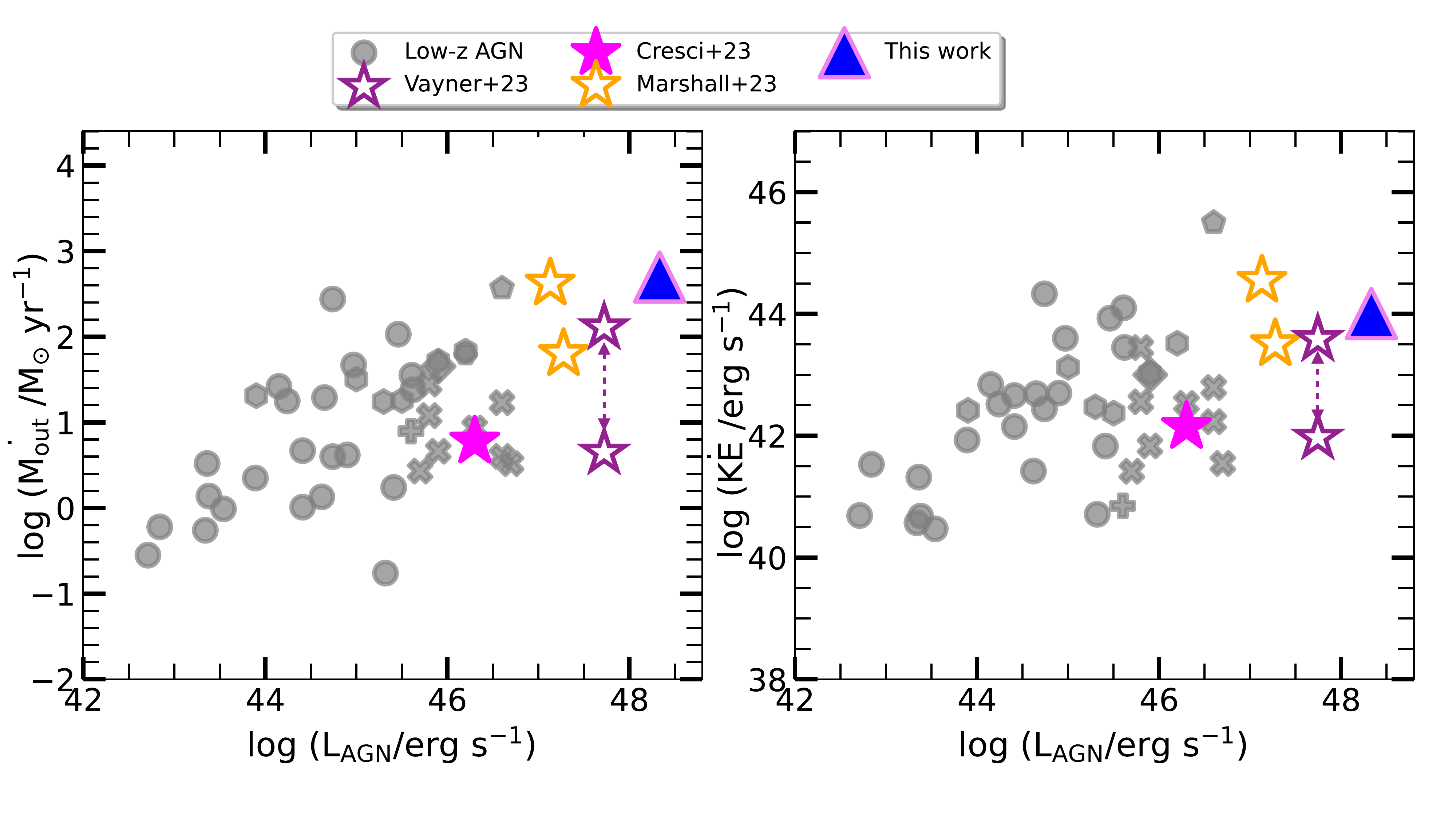}
    \caption{[Left panel] Globally integrated mass outflow rate vs. AGN bolometric luminosity for low-redshift quasars, reported in the literature using a variety of IFU/long-slit spectroscopy \citep{revalski18, nesvadba06, oliveira21, brusa16, revalski21, kakkad22}. The colored stars indicate estimates from JWST/IFU observations of radio-quiet quasars at high redshifts \citep{cresci23, veilleux23, marshall23, vayner23}. The unfilled stars represent measurements where the electron density estimates are assumed, and not directly measured due to non-detection of [SII] 6717, 32 lines. Hence the measured values are likely over/under-represented. For \cite{vayner23}, which reported outflow rates of a z = 3 radio-quiet quasar, we show two measurements by two stars (joined by a dashed arrowhead sign in between). These two measurements correspond to outflow rates at 1 Kpc and 5 Kpc radial distance respectively, based on their reported radial profiles. We do not use their summed ``global'' outflow rates for reasons discussed in the text. Our radio galaxy source TNJ1338 is shown with a blue triangle. TNJ1338 occupies the highest end of the distribution, with the highest outflow rates and the strongest AGN bolometric luminosity observed; but is overall consistent if we extrapolate the trend visible in radio-quiet quasars at high-z. [Right panel]  Kinetic outflow power vs. AGN bolometric luminosity for literature compiled objects (gray symbols and colored stars) compared to our source (triangle). The kinetic outflow power and the bolometric AGN luminosity recorded in our source are some of the strongest as measured at similar redshifts in radio-quiet quasars. We note that the outflow rates for TNJ1338 do not include corrections for projection effects (\S 4.3) and are hence lower limits. }
    \label{fig:qso_comparison}
\end{figure*}

 Our study presents the first detailed, spatially resolved view of a $z>4$ radio galaxy hosting outflow. Here, we compare the outflow properties of our source with four other high redshift (z $>$1.5) Type 1 ``unobscured'' quasar host galaxies recently studied with JWST/NIRspec IFU data, as reported in \cite{vayner23, cresci23, veilleux23, marshall23}. 

The outflow velocity and [OIII]-line widths for TNJ1338, characterized by Moment 1 and $\rm W_{50}$ values, exceed 1200-1500 $\rm km \ s^{-1}$, which is pretty high. Fig.~\ref{fig:kin_hist} shows the distribution of $\rm W_{80}$ measurements for our radio galaxy TNJ1338 (in black), compared with five other Type 1 quasars (in yellow, magenta, salmon, blue, and green) published with JWST NIRSpec IFU. For published sources who do not report $\rm W_{80}$ measurements, we assume $\rm W_{80}$ is approximately equal to twice the 2nd-moment values. We see that the kinematics for our system is the most extreme, with $\rm W_{80}$ reaching 2500-3000 $\rm km \ s^{-1}$ for some locations in the galaxy, with typical values ranging between 800-2900 $\rm km \ s^{-1}$. Although, the high $\rm W_{80}$ measurements are not unexpected for luminous AGNs, as shown in various studies \citep{perna17, coatman19, villar-martin21, bischetti17, temple19}, the high velocities over an extended region observed in TNJ1338 is still pretty spectacular.

 For TNJ1338, the total mass outflow rate ($\rm \dot{M_{out}} \sim 500 \ M_{\odot} \ yr^{-1} $) is roughly 
 similar to the star formation rate (SFR) estimated from the host galaxy region by \cite{duncan23}, which is $\sim$ 490 $\rm M_{\odot} \ yr^{-1}$. This SFR estimate is in agreement with the estimate of $\sim \rm 461 \ M_{\odot} \ yr^{-1}$ derived from the infrared continuum \citep{falkendal19}. Note both these SFR estimates are possibly upper limits due to significant contamination from AGN photoionization. The lower limit on the mass loading factor $\rm \eta = \dot{M}/SFR$ is thus $\sim$ 1.01.  The high mass outflow rate and mass loading imply that the outflow can efficiently displace a significant amount of gas to large distances. Unlike our system, the Type 1 QSO X1D2028 observed recently with JWST/NIRSpec as part of the JWST ERS-1335 program, shows a much lower mass loading factor $\eta \sim $ 0.04 \citep{veilleux23} and very low mass outflow rate \citep[$\sim \rm 6 \ M_{\odot} \ yr^{-1}$;][]{cresci23} in spite of hosting a powerful quasar. It is worth noting that this same quasar was previously reported to host a much higher mass outflow rate ($\sim \rm 100  \ M_{\odot} \ yr^{-1}$) using ground-based spectral analyses and with a generic assumption of electron density. With the improved JWST spectra, upgraded velocity values from better-resolved kinematics studies, and better estimates of density from [SII] 6717, 6731 lines, the mass outflow rate turned out to be $\sim $ 15 times lower. 
 
 The two Type 1 QSOs from \cite{marshall23} reported mass loading factors greater than unity (1.6 and 2). However, these mass outflow estimates are possibly highly uncertain, given that electron densities were not directly measured from [SII] 6717, 6731 line ratios owing to their non-detection, and were assumed to be a generic value. This can significantly affect the mass outflow rate estimates (63 and 420 $\rm M_{\odot}/yr$ ). \cite{vayner23} studied a powerful quasar-driven outflow at z = 3 and measured a mass outflow rate that declined steeply with radius, falling from $\rm \dot{M_{out}} = 100 \ M_{\odot} yr^{-1}$ at a radius of 1 kpc to $\sim \rm 3 \ M_{\odot} yr^{-1}$ at 5 kpc. In most of these cases, the outflow velocities are also tentative, since they assume the highest outflow velocity in the energetics calculation, which might not be representative of the major part of the outflowing region. 
 
Fig.~\ref{fig:qso_comparison} (left panel) shows the mass outflow rates ($\rm \dot{M_{out}}$) vs. bolometric AGN luminosity ($\rm L_{AGN}$) for our source (triangle) compared with published JWST Type 1 QSOs at high-z with spatially-resolved outflows \citep[stars;][]{marshall23, vayner23, cresci23} and a compilation of low redshift AGNs with similar outflow signatures from the literature \citep[gray symbols;][]{nesvadba06, brusa16, revalski18, shimizu19, revalski21, oliveira21, kakkad22}. We choose only those low-z studies that use directly measured electron density and outflow velocity in the calculation. Of the 4 high-z Type 1 QSOs shown, only one has density measurements and that is shown with a filled star symbol. The rest are shown with unfilled star symbols. We show the \cite{vayner23} outflow rate measurements as two separate points (purple stars) connected by a dashed arrowhead line. These represent the outflow rates reported at 1 kpc and 5 kpc radial distances respectively, as reported from their radial profiles. However, we do not sum up those estimates to derive a single measurement, as summing the contribution from different radii leads to an erroneous global outflow rate due to the specific way they calculate outflow rates.  
Our object lies at the highest end of both $\rm \dot{M_{out}}$ and $\rm L_{AGN}$ values, but is overall consistent with the linear correlation observed between these two quantities at low redshift. A similar observation can be made in the kinetic power vs. AGN luminosity diagram, shown in the right panel. 

Our source shows a total kinetic power of 1.01$\times \rm 10^{44} \ erg \ s^{-1}$. 
Various feedback models and simulations predict that for feedback to launch a blastwave sufficiently powerful to entrain gas and to inject adequate energy into the ISM, the kinetic power of the outflow needs to be about 5-7\% of the AGN bolometric luminosity. On the other hand, a `two-stage' model was proposed by \cite{hopkins10} that stated that initial feedback from the central quasar needs to only initiate a moderate wind in the low-density hot gas, which reduces the required energy budget for feedback by an order of magnitude. In this model, kinetic power needs to be a minimum of $\sim 0.5$\% of the AGN bolometric luminosity. Our source has a kinetic power $\rm \dot{KE}_{outflow} \sim 0.1$\% of $ \rm  L_{AGN}$, 
indicating that the outflow itself may not be able to deposit an enormous amount of energy which is capable of altering the surrounding ISM conditions. However, we should note two things. First, this kinetic power is only a lower limit, since it does not include corrections for projection effects (section 4.3). Second, the NIRSpec data only probes the warm ionized gas phase of the outflow; the other phases of the outflow -- like the hot phase or the cold molecular phase -- may carry more energetic outflows and the total kinetic power (summing all phases) may well exceed the 0.5\% threshold predicted by the two-stage feedback model.

\subsection{Radio-jet powered feedback} \label{subsec:jet-feedback}

The VLA images of TNJ1338 (Fig.~\ref{fig:tnj_RGB}) show 
 two blobs located to the north and south in opposite directions from the nuclear region, indicating a bi-polar radio jet \citep{pentericci00}. 
We have presented multiple pieces of evidence that the energy deposited by these powerful radio jets is responsible for the extraordinary kinematics exhibited by the ionized gas (traced by [OIII]$\lambda$5007, H$\alpha$, etc) over the region extending from roughly 3 kpc to 10 kpc to the north of the nucleus. If this bright and kinematically-disturbed nebular (N) region represents the site of a collision of a jet with a massive gas cloud, it would naturally explain why the northern radio lobe is about four times brighter than the southern lobe, and about three times closer to the nucleus. This is also in agreement with numerical simulations of jets modeled in a clumpy ISM \citep{dutta24}, which have shown that the radio jet undergoes deceleration when the jet head is obstructed by dense clouds, and results in a strong shock that creates radio hotspots. 
We also showed that the kinematics in this region are far more extreme than what has been observed in JWST NIRSpec IFU data for outflows in radio-quiet QSOs with bolometric luminosities similar to TNJ1338 (Fig.~\ref{fig:kin_hist}).

\begin{figure*}
    \centering
    \includegraphics[width=\textwidth]{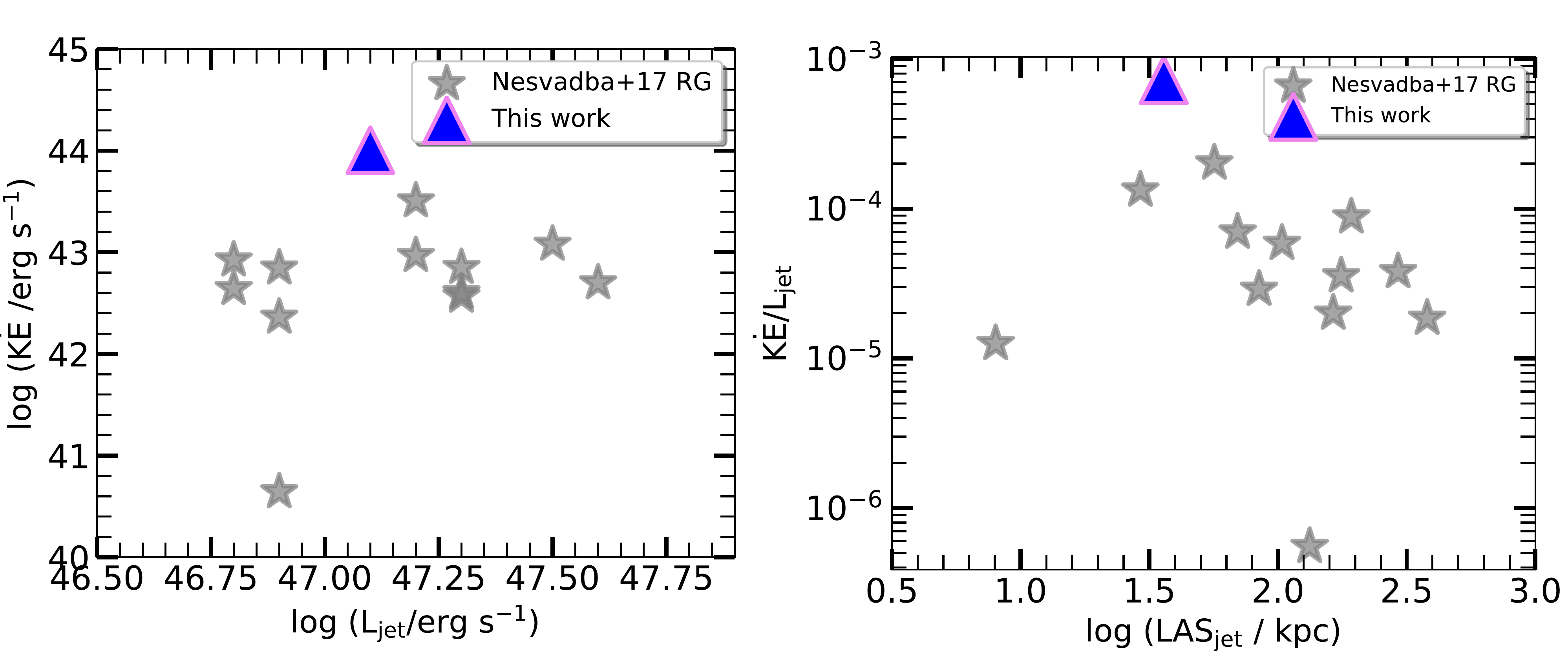}
    \caption{[Left panel] Kinetic power of the outflows vs. the radio jet mechanical energy for high-z (median z $\sim$ 2.5) radio galaxy sample reported in \cite{nesvadba17} (gray stars), and our source for this work TNJ1338 (blue triangle). [Right panel]  Radio of kinetic power and jet mechanical energy plotted against radio size (in kpc) for the same objects. TNJ1338 has the highest kinetic power of outflows, with $\rm \dot{KE}_{outflow}$ about an order of magnitude larger than the median value for the other radio galaxies in the N17 sample. No corrections for projection effects (see \S\ref{subsec:o3_energy}) have been made. The radio size of TNJ1338 is also relatively small, suggesting that the transfer of kinetic energy from the jet to the gas is facilitated by the higher gas densities expected at smaller radii.  }
    \label{fig:kedot_ljet}
\end{figure*}


In this section, we will quantify the nature of the jet/ISM interaction that we are witnessing. 
We begin with a comparison of the energetics of the outflow in TNJ1338 to those of the other high-z radio galaxies investigated by \cite{nesvadba17}[N17] using ground-based NIR IFU data. TNJ1338 has a similar radio luminosity (and inferred jet kinetic energy flux) with most of the N17 sample but differs in terms of its redshift (z = 4.104 vs. the N17 median of z = 2.5) and the compact size of the radio source ($\sim$ 36 kpc {\it vs.} the N17 median of $\sim$ 100 kpc). In fact, TNJ1338 is in the N17 sample, but its optical emission-line region was spatially unresolved from ground-based observations. With the JWST/NIRSpec revealing extended ionized gas component in TNJ1338, we aim to make an `apples to apples' comparison of the energetics of this emission-line gas with the rest of the high-z radio galaxies from the N17 sample. N17 computed two estimates of  $\rm \dot{KE}_{outflow}$, but these used entirely different approaches than ours. Hence, we do not use those for comparison. We have used the data tabulated in N17 to measure $\rm \dot{KE}_{outflow}$ ourselves for the other high-z radio galaxies in exactly the same way as we calculated for TNJ1338, as described in \S 4.3. 
The results are shown in Fig.~\ref{fig:kedot_ljet}.  In the left panel, we plot  $\rm \dot{KE}_{outflow}$ vs. $\rm L_{jet}$ for TNJ1338 (blue triangle) and N17 radio galaxy sample (gray stars). There is no evident correlation, but this may be due in part to the small range spanned in $\rm L_{jet}$ (a factor of $\sim$ 6). TNJ1338 is an outlier, with $\rm \dot{KE}_{outflow}$ about an order of magnitude larger than the median value for the other radio galaxies.

To investigate this further, in the right panel, we plot the ratio of $\rm \dot{KE}_{outflow}$/$\rm L_{jet}$ vs. the size of the radio source. There is a trend for the former ratio to increase as the radio source size decreases, with TNJ1338 lying at one end of the distribution. This suggests that the transfer of kinetic energy from the jet to the gas is facilitated by the higher gas densities expected at smaller radii. However, it is important to note that in the case of TNJ1338, the kinetic energy flux in the gas is less than 0.1\% of the jet kinetic energy flux. Even correcting for projection effects (\S 4.3) the kinetic energy flux will be less than 1\% of the jet kinetic energy flux.

\begin{figure*}
    \centering
    \includegraphics[width=0.97\textwidth]{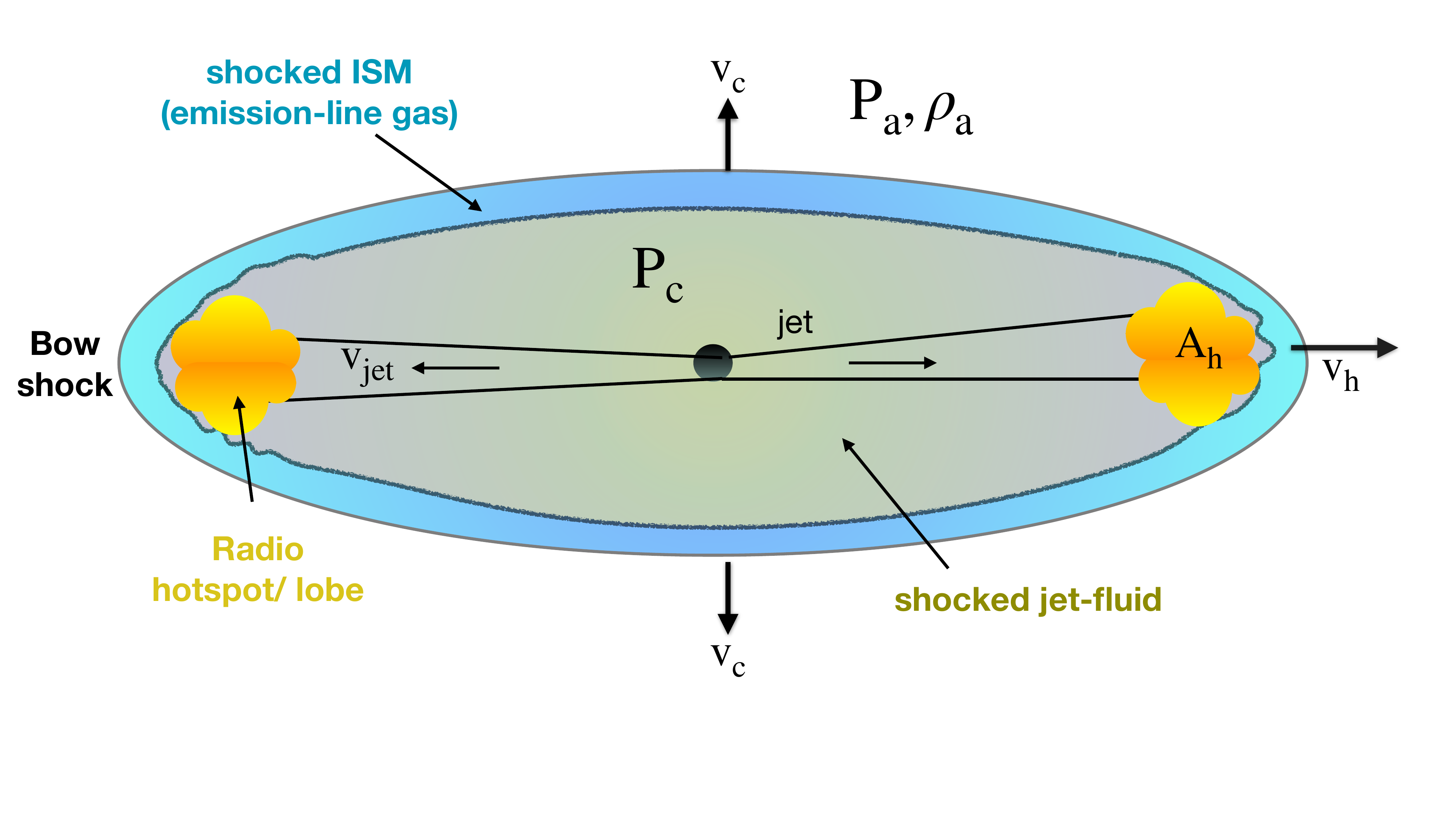}
    \caption{ Schematic diagram to summarize the proposed scenario in TNJ1338, inspired by the model proposed by \cite{begelman89}. The radio jet, traveling with a speed of $\rm v_{jet}$, terminates with the formation of radio lobes/hot spots, with a cross-sectional area $\rm A_h$. While the jets plow through the surface, they are enveloped in a ``cocoon'' consisting of shocked jet-fluid material, which is wrapped in a layer of shocked interstellar medium ambient gas. This thin layer of shocked ISM emission-line gas is possibly visible in the JWST/NIRSpec IFU data as extended [OIII] emission, spatially aligned with the radio lobes. $\rm P_c$, and $\rm P_a$ indicate the pressure inside and outside the cocoon respectively. If the cocoon is overpressured, i.e. $\rm P_c > P_a$, it drives a shock laterally -- thus expanding the cocoon sideways with a velocity $\rm v_c$, and with a velocity $\rm v_{h}$ in the orthogonal direction. Utilizing the dynamical timescale $\sim$ 3 Myr derived from this simplistic picture, we compute the total kinetic energy supplied by the radio jet, and find that the kinetic energy is transferred very inefficiently from the jets to the emission-line gas ($\rm KE_{gas}/KE_{jet} \sim 10^{-3}$), consistent with the right panel of Fig.~\ref{fig:kedot_ljet}.   }
    \label{fig:cartoon}
\end{figure*}

We adopted our method of computing outflow rates in \S 4.3 to be as similar as possible to those in previous JWST NIRSpec IFU investigations of outflows of ionized gas in high-z radio-quiet QSOs, so that TNJ1338 could be placed into this context (see Fig.~\ref{fig:kedot_ljet}). This method implicitly assumes that the observed gas has traveled from the nucleus to its current location at its currently observed velocity (that is, mass and kinetic energy [KE] are converted into fluxes using a crossing-time of $t_{cross} = R/v_{out}$). While this is a natural choice, it may be inappropriate for the situation seen in TNJ1338 (or other jet-driven outflows). In this case, the appropriate timescale for the outflow would be the lifetime of the radio source ($t_{RS}$). This can be significantly shorter than $t_{cross}$. In this case, the observed gas has only been moving at $v_{out}$ for a timescale $t_{RS}$, meaning it has moved only a relatively small distance from its initial location.
We can quantify this picture in the case of TNJ1338. To estimate the radio source lifetime $t_{RS}$, we need to take the distance of the northern radio lobe from the nucleus (8.8 kpc) and divide it by an estimate of the outward velocity of the radio lobe ($v_{RL}$). We will do this in two ways: one empirical and one based on simple analytic theory. We begin with the latter since it provides a simple conceptual framework for interpreting the data.

\cite{begelman89} presented a simple theoretical model for how a radio jet would interact with ambient gas. In this model, the jet terminates in a strong shock, producing intense radio emission from a “hot spot”. The immensely over-pressured shocked-jet fluid then expands laterally and back-flows to create an expanding ellipsoidal “cocoon” that pushes the gas and drives outflows into the surrounding gas. This agrees with high-resolution hydrodynamic jet simulations which have found similar wide cocoons expanding through the low-density diffuse phase \citep{dutta24}. Fig.~\ref{fig:cartoon} summarizes this picture. In this scenario, they showed that the velocity at which the hot-spot advances is given by:
\begin{equation}
\rm v_h = (\frac{L_{jet}}{(A_h v_{jet} \rho_a)})^{1/2}
\end{equation}
Here, $\rm L_{jet}$ is the rate the jet transports kinetic energy, $A_h$ is the cross-sectional area of the hotspot, $v_{jet}$ is the jet velocity \citep[which is take as $c$ in ][]{begelman89}, and $\rho_a$ is the mean ambient gas density. These are marked in Fig.~\ref{fig:cartoon}. For TNJ1338 the estimated value for the  $\rm L_{jet}$ is $7 \times 10^{46}$ $\rm erg \ s^{-1}$ for the northern jet alone \citep{nesvadba17}. Based on the lateral extent of the northern radio lobe in Fig.~\ref{fig:tnj_RGB}  \& \ref{fig:kin_hist}, we estimate $A_h = 3 \times 10^{43} \rm \ cm^2$. We take $\rm \rho_a = 1.6 \times 10^{-24} \ cm^{-3}$, so that the total gas mass is $10^{11}$ M$_{\odot}$ interior to a radius of 10 kpc (similar to the galaxy stellar mass – Table 1). With these values, the predicted value is $v_h \sim$ 2200 $\rm km \ s^{-1}$, corresponding to a radio source lifetime $t_{RS} \sim 3.9$ Myr. 

Using this conceptual framework, we can get an independent empirical estimate of the radio source lifetime using the results on the double-peaked emission-line profiles lying along the jet axis (Fig.~\ref{fig:o3profile}). In the \cite{begelman89} model these line profiles trace the lateral expansion velocity of the cocoon (with one line each from the front and back sides). The mean peak separation, i.e., $\rm v_{c} \sim$ 900 $\rm km \ s^{-1}$, and the lateral extent of the region is about 3 kpc. Dividing this size by this velocity implies a dynamical age of about 3.3 Myr, which is close to the theoretical estimate above. In comparison, $t_{cross}$ for the emission-line region (see above) is about 10 Myr.
In this picture in which the gas is very rapidly accelerated for only a short time, it makes more sense to compare the total KE in the gas to the total KE transported by the jets (rather than comparing the $\rm \dot{KE}$). For a radio source lifetime of 3.3 Myr and a two-sided jet kinetic energy flux $\rm L_{jet}$ of 1.4 $\times 10^{47}$ $\rm erg \ s^{-1}$, the total amount of injected energy is $E_{jet} \sim 1.4 \times 10^{61}$ erg. This can be compared to the total KE measured in the outflowing ionized gas $\rm \dot{KE}_{outflow} \sim $ 3 $\times 10^{58}$ erg (or 1.2 $\times 10^{59}$ erg for a median correction for projection effects - see section 4.3). This implies that KE is transferred very inefficiently from the jets to the emission-line gas (consistent with the results in Fig.~\ref{fig:kedot_ljet} above).

Next, we use equation (6) in \cite{begelman89}, to estimate the pressure in the northern cocoon, using the values above for $L_{jet}$  and $A_h$, a cocoon length of 8.8 kpc and a cocoon half width of 1.5 kpc. This yields $P_c = 3.4 \times 10^{-7}$ \ dyne \ cm$^{-2}$.   This is roughly $10^2$ times larger than the pressure measured using the [SII] derived densities and an assumed temperature for photoionized gas of $\sim 10^4$ K \citep{osterbrock06}.  This implies that the gas we see does not lie inside the cocoon, but instead represents ambient lower-pressure gas surrounding (and being accelerated by) the cocoon.

The result that the emission-line gas carries only a small fraction of the jet kinetic energy should not be surprising. If the cocoon is traveling outward at $\rm v_{h} \sim$ 2200 $\rm km \ s^{-1}$ and expanding laterally at $\rm v_{c} \sim$ 500 $\rm km \ s^{-1}$ (see above), the shocks that these motions drive into the ambient gas correspond to post-shock temperatures of $\sim$ 80 million and 4 million K respectively. Thus, the bulk of the deposited energy would be in the form of hot X-ray-emitting gas. In fact, as noted in the Introduction, TNJ1338 has been detected in X-rays \citep{smail13}. The X-ray emission is aligned with the radio axis and is extended by about 30 kpc, but no detailed spatial coincidence exists between the X-ray and radio emission. 
We suggest that the X-ray emission may arise in shocks driven by the expanding cocoon. The X-ray luminosity is 3 $\times 10^{44}$ $\rm erg \ s^{-1}$ which, while substantial, is far less than the KE flux carried by the jets. This would imply that the radiative cooling time of the hot gas is much longer than the radio source's lifetime.

\section{Conclusions}
In this paper, we present the JWST NIRSpec/IFu observations (JWST GO 1964: co-PIs Overzier, Saxena) of TNJ1338, a radio galaxy at z = 4.104, hosting an obscured quasar at the nucleus. This source is a double-lobed radio galaxy as revealed from existing VLA radio observations, with the northern lobe $\sim$ 4 times brighter than the southern lobe, and about three times closer to the nucleus. This bright radio lobe spatially overlaps with an extended ionized gas nebula which shows some of the fastest-moving outflows compared to other unobscured quasar populations recently studied with JWST. We map the location of these fast-moving outflows, characterize the spatially resolved outflow kinematics and energetics at kpc-scale resolution using rest-frame optical emission lines using JWST/NIRSpec IFU, and construct a simplistic model to explain the radio jet driven feedback possibly at play. 
The following are the main conclusions: 

\begin{enumerate}
    \item We detect ionized outflow with velocities exceeding 900 $\rm km \ s^{-1}$ and broad line profiles with line widths exceeding 1200 $\rm km \ s^{-1}$ (mean $\sim \rm W_{50} \sim $ 800 $\rm km \ s^{-1}$, mean $\rm W_{80} \sim $ 1600 $\rm km \ s^{-1}$) about $\sim$ 5 kpc away from the central nucleus of the galaxy. The outflowing gas extends up to $\rm \sim 15 $ kpc in projection from the center, roughly co-spatial with the bright northern radio lobe.

    \item A massive amount of ionized gas of mass $\sim \rm 5 \times 10^{9} \ M_{\odot}$ is detected, with a total integrated mass outflow rate of 497 $\rm M_{\odot} \ yr^{-1}$. The total kinetic power deposited by the outflows is $\rm \sim 1 \times 10^{44} \ erg \ s^{-1}$. We present carefully measured estimates of spatially resolved mass outflow rates and kinetic energy for the first time for a z$>$4 AGN host galaxy.

    \item The extended ionized gas nebula detected is likely formed due to photoionization by the obscured quasar residing in the host galaxy. However, the powerful radio jets are responsible for the large-scale outflows and the extraordinary kinematics exhibited by the ionized gas. Our hypothesis is that the radio jet terminates in a
strong shock, producing radio ``hot spot/lobes''. The immensely over-pressured shocked jet fluid then expands laterally and back-flows to create an expanding ellipsoidal “cocoon” that entrains/ accelerates the
surrounding gas outwards -- creating this high-velocity gas. This scenario explains the kinematics and the spatial alignment between the northern radio lobe and the optical emission line gas seen in TNJ1338. 

    \item The total kinetic energy injected by the radio jet is $\rm \sim 1\times 10^{61} \ erg $, which is roughly 3 orders of magnitude larger than the total kinetic energy measured in the outflowing ionized gas. This
implies that kinetic energy is transferred very inefficiently from the jets to the emission-line gas. This is unsurprising, since the expanding cocoons drive shocks corresponding to post-shock temperatures of a few million K. Hence the bulk of the deposited energy would be in the form of hot X-ray-emitting gas. 

\item Our source has a kinetic efficiency ($\rm \dot{KE}_{outflow}$/ $  L_{AGN}$) $\sim$ 0.1\%, 
indicating that the outflow itself may not be able to deposit an enormous amount of energy which is capable of altering the surrounding ISM conditions.

\end{enumerate}




\begin{acknowledgments}
NR and TH thank Anton Koekemoer for their help in the registration of the NIRSpec cube with the NIRCAM mosaics. KJD acknowledges support from the STFC through an Ernest Rutherford Fellowship (grant number ST/W003120/1). S.E.I.B. is supported by the Deutsche Forschungsgemeinschaft (DFG) through Emmy Noether grant number BO 5771/1-1.
 
\end{acknowledgments}

%



\software{Astropy \citep{astropy18},  
          Scipy \citep{scipy20}, 
          JWST STScI pipeline \citep{stsci}, 
          dynesty \citep{speagle20}
          }



\appendix

\bibliography{main}{}
\bibliographystyle{aasjournal}



\end{document}